\pdfoutput=1
\documentclass[10pt]{article}

\usepackage{amsmath}
\usepackage{amssymb}

\usepackage{graphicx}

\usepackage{cite}

\usepackage{color} 

\usepackage[labelfont=bf,labelsep=period,justification=raggedright]{caption}

\makeatletter
\renewcommand{\@biblabel}[1]{\quad#1.}
\makeatother

\date{}

\usepackage{hyperref}
\usepackage{setspace}
\newcommand{\degree}{\ensuremath{^\circ}}
\newcommand{\plusminus}{\ensuremath{\pm}}
\newcommand{\beginsupplement}{\setcounter{table}{0}\renewcommand{\thetable}{S\arabic{table}}\setcounter{figure}{0}\renewcommand{\thefigure}{S\arabic{figure}}}

\begin{document}
\begin{doublespacing}
\begin{flushleft}
{\Large
\textbf{\textit{Drosophila} embryogenesis scales uniformly across temperature in developmentally diverse species}
}
\\
Steven G. Kuntz$^{1,2\ast}$, 
Michael B. Eisen$^{1,2,3,4}$
\\
\bf{1} QB3 Institute for Quantitative Biosciences, University of California, Berkeley, California, United States of America
\\
\bf{2} Department of Molecular and Cell Biology, University of California, Berkeley, California, United States of America
\\
\bf{3} Howard Hughes Medical Institute, University of California, Berkeley, California, United States of America
\\
\bf{4} Department of Integrative Biology, University of California, Berkeley, California, United States of America
\\
$\ast$ E-mail: sgkuntz@berkeley.edu
\end{flushleft}

\section*{Abstract}
Temperature affects both the timing and outcome of animal development, but the detailed effects of temperature on the progress of early development have been poorly characterized. To determine the impact of temperature on the order and timing of events during \textit{Drosophila melanogaster} embryogenesis, we used time-lapse imaging to track the progress of embryos from shortly after egg laying through hatching at seven precisely maintained temperatures between 17.5\degree C and 32.5\degree C. We employed a combination of automated and manual annotation to determine when 36 milestones occurred in each embryo. \textit{D. melanogaster} embryogenesis takes $\sim$33 hours at 17.5\degree C, and accelerates with increasing temperature to a low of 16 hours at 27.5\degree C, above which embryogenesis slows slightly. Remarkably, while the total time of embryogenesis varies over two fold, the relative timing of events from cellularization through hatching is constant across temperatures. To further explore the relationship between temperature and embryogenesis, we expanded our analysis to cover ten additional \textit{Drosophila} species of varying climatic origins. Six of these species, like \textit{D. melanogaster}, are of tropical origin, and embryogenesis time at different temperatures was similar for them all. \textit{D. mojavensis}, a sub-tropical fly, develops slower than the tropical species at lower temperatures, while \textit{D. virilis}, a temperate fly, exhibits slower development at all temperatures. The alpine sister species \textit{D. persimilis} and \textit{D. pseudoobscura} develop as rapidly as tropical flies at cooler temperatures, but exhibit diminished acceleration above 22.5\degree C and have drastically slowed development by 30\degree C. Despite ranging from 13 hours for \textit{D. erecta} at 30\degree C to 46 hours for \textit{D. virilis} at 17.5\degree C, the relative timing of events from cellularization through hatching is constant across all species and temperatures examined here, suggesting the existence of a previously unrecognized timer controlling the progress of embryogenesis that has been tuned by natural selection as each species diverges.

\section*{Author Summary}
Temperature profoundly impacts the rate of development of ``cold-blooded" animals, which proceeds far faster when it is warm. There is, however, no universal relationship. Closely related species can develop at markedly different speeds at the same temperature. This creates a major challenge when comparing development among species, as it is unclear whether they should be compared at the same temperature or under different conditions to maintain the same developmental rate. Facing this challenge while working with flies  (\textit{Drosophila} species), we found there was little data to inform this decision. So, using time-lapse imaging, precise temperature-control, and computational and manual video-analysis, we tracked the complex process of embryogenesis in 11 species at seven different temperatures. There was over a three-fold difference in developmental rate between the fastest species at its fastest temperature and the slowest species at its slowest temperature. However, our finding that the timing of events within development all scaled uniformly across species and temperatures astonished us. This is good news for developmental biologists, since we can induce species to develop nearly identically by growing them at different temperatures. But it also means flies must possess some unknown clock-like molecular mechanism driving embryogenesis forward. 
 
\section*{Introduction}
It has long been known that \textit{Drosophila}, like most poikilotherms, develops faster at higher temperatures, with embryonic \cite{Powsner:1935}, larval \cite{Powsner:1935, James:1995tx}, and pupal stages \cite{James:1997tr, Yamamoto1982}, as well as total lifespan \cite{Lillie:1897, Loeb1916} showing similar logarithmic trends. While genetics, ecology, and evolution of this trait have been investigated for over a century \cite{James:1995tx, Partridge1994, Kim:2000ui, Tantawy1963, Strataman1998, Morin1997evo, Montchamp-Moreau1983, Hoffmann:2001ut, Hoffmann:2007hr, Rako:2007gm, Gibert:2001, MarkowAndOGrady2005}, the effects of temperature on the order and relative timing of developmental events, especially within embryogenesis, are poorly understood.

We became interested in the relationship between species, temperature, and the cadence of embryogenesis for practical reasons. Several years ago, we initiated experiments looking at the genome-wide binding of transcription factors in the embryos of divergent \textit{Drosophila} species: \textit{D. melanogaster}, \textit{D. pseudoobscura}, and \textit{D. virilis}. With transcription factor binding a highly dynamic process, we tried to match both the conditions (especially temperature, which we believed would affect transcription factor binding) in which embryos were collected and the developmental stages we analyzed. However, our initial attempts to collect \textit{D. pseudoobscura} embryos at 25\degree C \textemdash the temperature at which we collect \textit{D. melanogaster} \textemdash were unsuccessful, with large numbers of embryos failing to develop, likely a consequence of \textit{D. pseudoobscura}'s alpine origin. While \textit{D. virilis} lays readily at 25\degree C, we found that their embryos develop more slowly than \textit{D. melanogaster}, complicating the collection of developmental stage-matched samples. 

Having encountered such challenges with just three species, and planning to expand to many more, we were faced with several important questions. Given that embryogenesis occurs at different rates in different species \cite{Kim:2000ui, Markow:2009}, how should we time collections to get the same mix of stages we get from our standard 2.5 \textendash 3.5 hour collections in \textit{D. melanogaster}, or any other stage we study in the future? Is it better to compare embryos collected at the same temperature even if it is not optimal for, or even excludes, some species; or, should we collect embryos from each species at their optimal temperature, if such a thing exists? Should we select a temperature for each species so that they all develop with a similar velocity? Or should we find a set of species that develop at the same speed at a common temperature? And even if we could match the overall rate of development, would heterochronic effects mean that we could not get an identical mix of stages? 

We found a woeful lack in the kind of data needed to answer these questions. Powsner precisely measured the effect of temperature on the total duration of embryogenesis in \textit{D. melanogaster} \cite{Powsner:1935}, and Markow made similar measurements for other \textit{Drosophila} species at a fixed temperature (24\degree C) \cite{Markow:2009}, but the precise timing of events within embryogenesis had been described only for \textit{D. melanogaster} at 25\degree C \cite{Campos-Ortega1985, Foe1993}. 

\begin{figure}[!ht]
\begin{center}
\includegraphics[width = 5.5in]{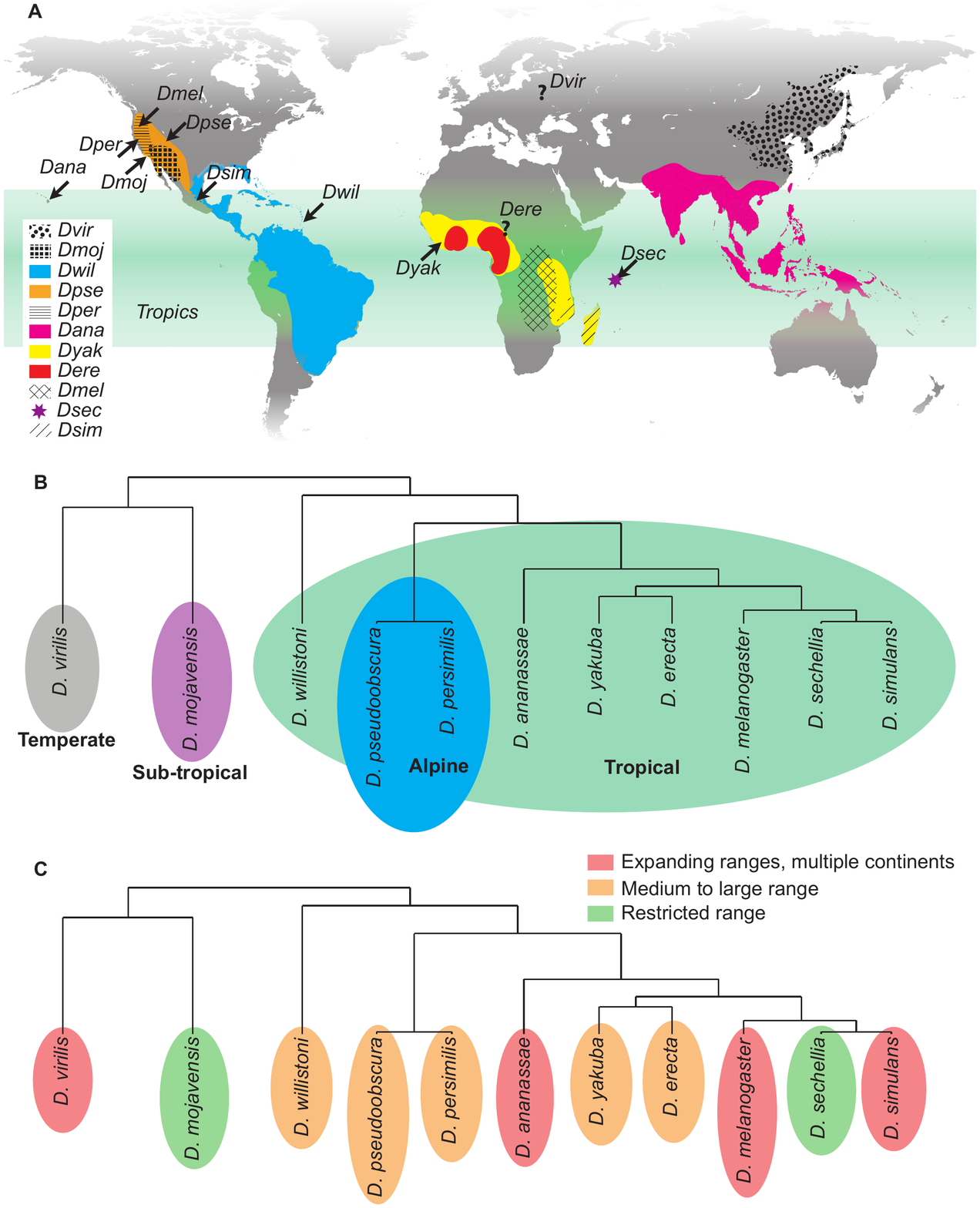}
\end{center}
\caption{\textbf{Geographic and climatic origin and phylogeny of analyzed \textit{Drosophila} species} (A) Ancestral ranges are shown for each species \cite{Lachaise:2004tj, MarkowAndOGrady2005, Patterson:1952}. While \textit{D. melanogaster} and \textit{D. simulans} are now cosmopolitan and \textit{D. ananassae} is expanding in the tropics (green), their presumed ancestral ranges are shown. \textit{D. virilis} is holarctic (gray) and restricted from the tropics, with a poor understanding of its ancestral range. Other species are more or less found in their native ranges, covering a variety of climates. Sites of collection are noted by arrows. (B) The phylogeny of the sequenced \textit{Drosophila} species. Many of the tropical species are closely related, though \textit{D. willistoni} serves as a tropical out-group compared to the melanogaster and obscura groups. Branch lengths are based on evolutionary divergence times \cite{Granzotto:2009}. (C) Range sizes vary considerably between the species.}
\label{fig:MapAndPhylogeneticTree}
\end{figure}

The work described here was born to address this deficiency. We used a combination of precise temperature control, time-lapse imaging, and careful annotation to catalog the effects of a wide range of temperatures on embryonic development in 11 \textit{Drosophila} species from diverse climates. We focused on species with published genome sequences \cite{Clark2007} (Table \ref{tab:SpeciesList}), as these are now preferentially used for comparative and evolutionary studies. Of the species we studied \textit{D. melanogaster}, \textit{D. ananassae}, \textit{D. erecta}, \textit{D. sechellia}, \textit{D. simulans}, \textit{D. willistoni}, and \textit{D. yakuba} are all native to the tropics, though \textit{D. melanogaster}, \textit{D. ananassae}, and \textit{D. simulans} have spread recently to become increasingly cosmopolitan \cite{MarkowAndOGrady2005}. \textit{D. mojavensis} is a sub-tropical species, while \textit{D. virilis} is a temperate species that has become holarctic and \textit{D. persimilis} and \textit{D. pseudoobscura} are alpine species (Figure \ref{fig:MapAndPhylogeneticTree}A).

\section*{Results}
\subsection*{Time-lapse imaging tracks major morphological events}

We used automated, time-lapse imaging to track the development of embryos held at a constant and precise temperature from early embryogenesis (pre-cellularization) to hatching. We maintained the temperature at \plusminus 0.1\degree C using thermoelectric Peltier heat pumps. Different sets of embryos were analyzed at temperatures ranging from 17.5\degree C to 32.5\degree C, in 2.5\degree C increments. Images were taken every one to five minutes, depending on the total time of development. A minimum of four embryos from each species were imaged for each temperature, for a total of 77 conditions. In total, time-lapse image series were collected and analyzed from over 1000 individual embryos.

\begin{figure}[h]
\includegraphics[scale=0.70]{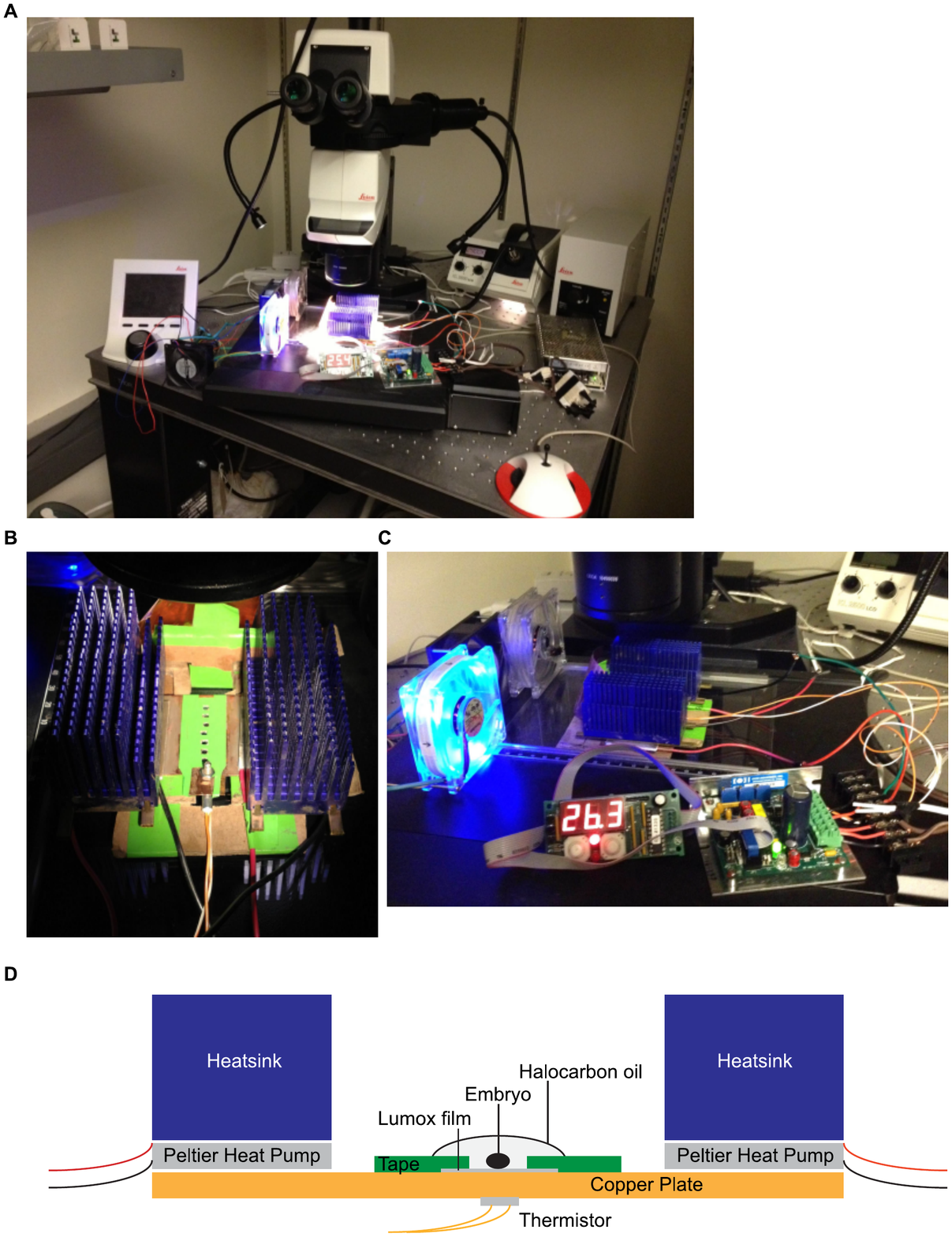}
\caption{\textbf{Microscopy imaging setup} (A) The imaging setup, showing the dissecting microscope with temperature control apparatus on the automated stage. (B) A close-up view of the temperature controlled platform flanked by heat-sinks (blue) that sit atop the Peltier thermoelectric controllers. In the center is a copper plate, with a thermister at the bottom to monitor plate temperature. The holes in the green masking tape line up with holes drilled through the copper plate and lined with a gas-permeable membrane. The masking tape helps retain the halocarbon oil. (C) A closer view of the setup. (D) A schematic of the setup demonstrates the temperature control and imaging apparatus in cross-section.}
\label{fig:ImagingSetup}
\end{figure}

We encountered, and solved, several challenges in designing the experimental setup, including providing the embryos with sufficient oxygen \cite{Foe:1983uv, Kam:1991vd} and humidity. We found that glass slides were problematic due to a lack of oxygenation and led to a $\sim$28\% increase in developmental time, so we instead employed an oxygen-permeable tissue culture membrane, mounted on a copper plate to maintain thermal conduction. At higher temperatures, we found that the embryos dehydrated, so humidifiers were used to increase ambient humidity. Detailed photos of the apparatus and descriptions can be found in Figure \ref{fig:ImagingSetup}.  

\begin{figure}[!ht]
\begin{center}
\includegraphics[width=5.5in]{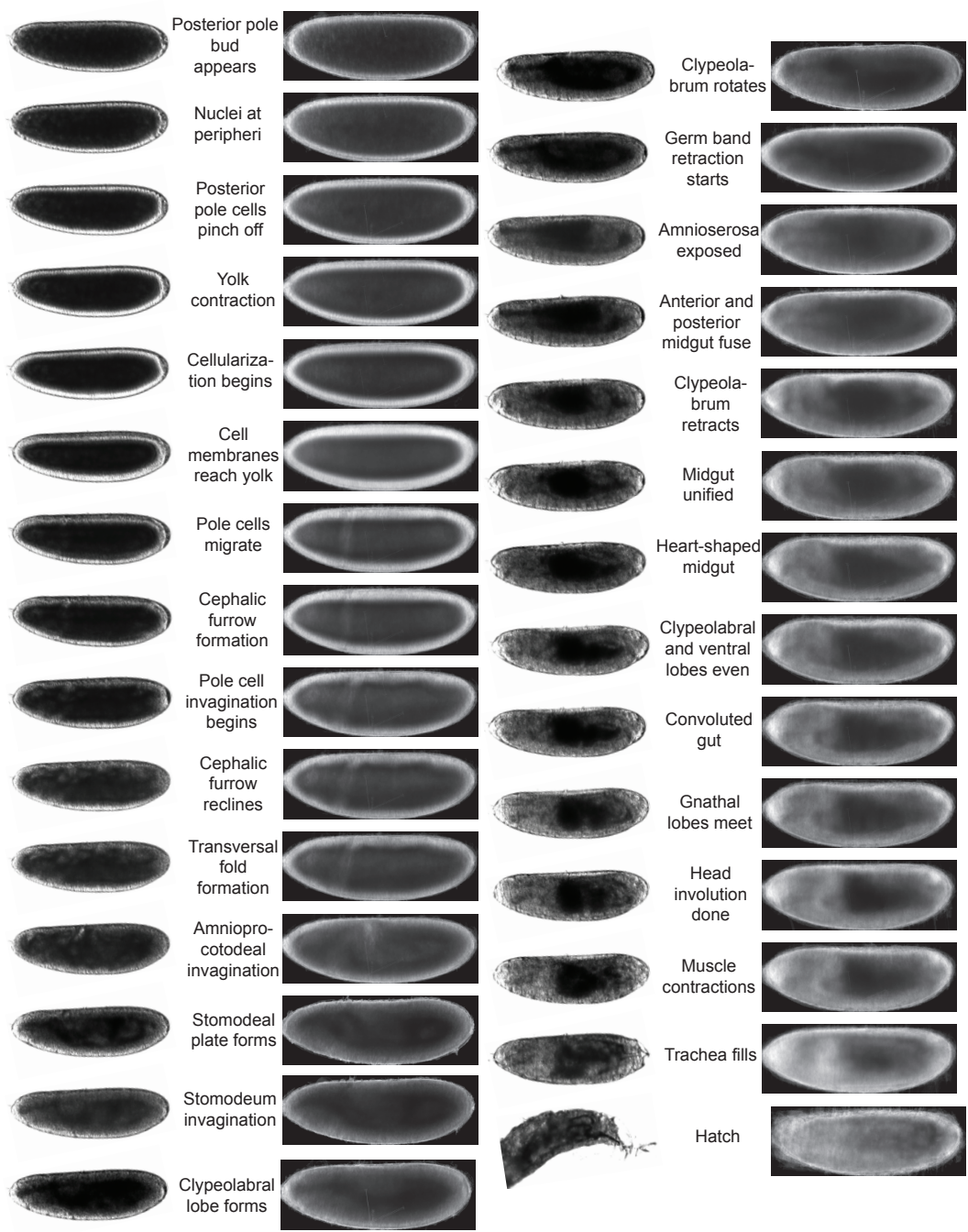}
\end{center}
\caption{\textbf{Developmental landmarks used in study} Many images of each  stage (examples on the left) were averaged to generate composite images (lateral view on the right) for each of the developmental stages, of which 29 are shown.}
\label{fig:TimeLapseImaging}
\end{figure}

\begin{figure}[h]
\includegraphics[scale=0.75]{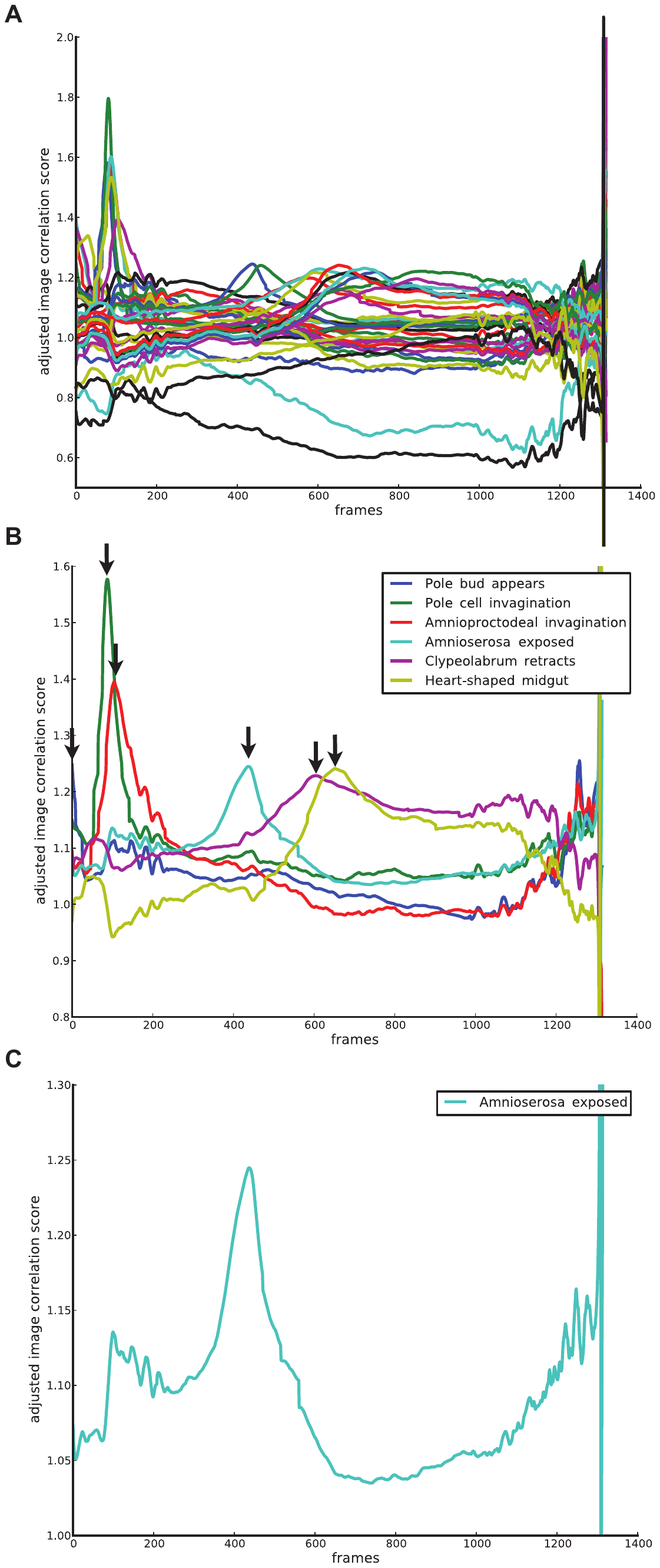}
\caption{\textbf{Events were predicted by computational analysis before manual verification} (A) For every time-lapse, each frame was correlated to each of the 34 composite images. (B) The running scores for 6 different events, with their maxima (black arrows) highlighted to reflect the estimated event time. (C) The time of amnioserosa exposure is estimated by the strong correlation at about 450 frames into the time-lapse.}
\label{fig:AutomatedEventPrediction}
\end{figure}

We used a series of simple computational transformations (implemented in Matlab) to orient each embryo, correct for shifting focus, and adjust the brightness and contrast of the images, creating a time-lapse movie for each embryo. We  manually examined images from ~60 time-lapse series in \textit{D. melanogaster} and identified 36 distinct developmental stages \cite{Campos-Ortega1985, Foe1993} that could be recognized in our movies (Table \ref{tab:EventList}, \url{http://www.youtube.com/watch?v=dYSrXK3o86I} and \url{http://www.youtube.com/watch?v=QKVmRy3dDR0} or ``\textit{D. melanogaster} with labelled stages" and ``\textit{D. melanogaster} with labelled stages at reduced framerate" in DOI:10.5061/dryad.s0p50''). Due to the volume of images collected, we implemented a semi-automated system to annotate our entire movie collection. Briefly, images from matching stages in manually annotated \textit{D. melanogaster} movies were averaged to generate composite reference images for each stage (Figure \ref{fig:TimeLapseImaging}). We then used a Matlab script to find the image-matrix correlation between each of these composite reference images to the images in each time-lapse to estimate the timing of each morphological stage via the local correlation maximum (Figure \ref{fig:AutomatedEventPrediction}A). 

\begin{figure}[h]
\includegraphics[scale=0.75]{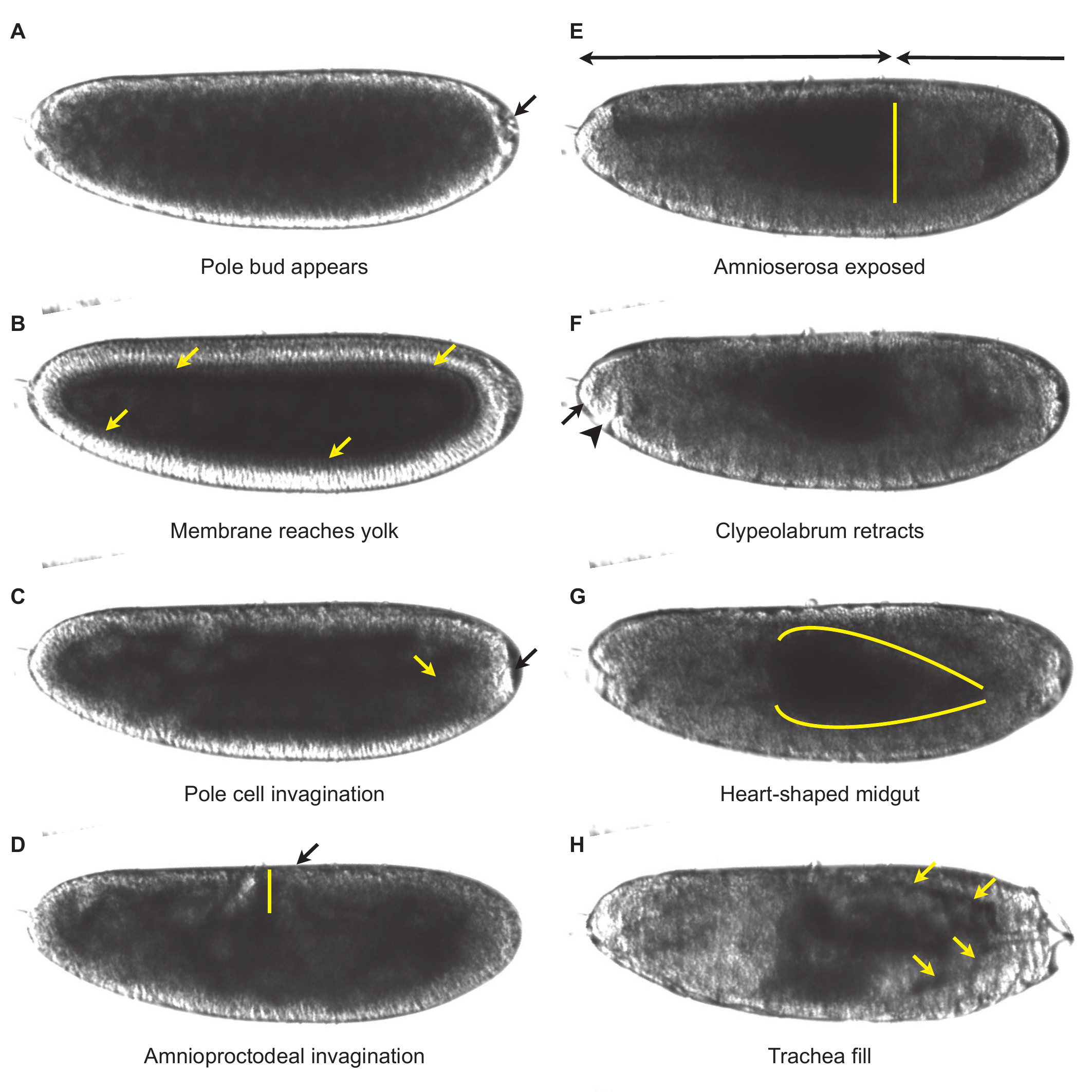}
\caption{\textbf{Identifying morphological stages} (A) `Pole bud appears' stage is identified by the first appearance of cells migrating into the posterior gap of the embryo (black arrow). (B) `Membrane reaches yolk' stage is identified by the converging of the leading edge of the invaginating cytoplasmic membrane on the dark yolk. (C) `Pole cell invagination' is identified by the completion of the fold (black arrow) that encapsulates the pole cells (yellow arow). (D) `Amnioproctodeal invagination' is identified by the point when the leading edge of the posterior invagination (black arrow) has covered $\sim$80\% of the distance to the leading edge of the cephalic furrow (vertical yellow line) and the pole cells have reached the interior of the embryo. (E) `Amnioserosa exposed' is identified by the point when the trailing edge of the germ band has retracted to the posterior 30\% of the embryo. (F) `Clypeolabrum retracts' is identified by the withdrawal of the ventral edge of the clypeolabrum (black arrow) from the gnathal buds and vitelline membrane to create a gap (black arrowhead). (G) `Heart-shaped midgut' is identified by the posterior elongation of the formerly spherical developing midgut and residual yolk (dark mass in the center of the embryo) to form a contiguous dark teardrop or heart-shaped mass (delimited with yellow lines).  (H) `Trachea fill' is identified by the rapid darkening of the trachea as they fill. The primary branches of the trachea run along the both the left and right dorsal sides, originating at the posterior of the embryo.}
\label{fig:DetailedStages}
\end{figure}

Of the 36 events, the eight most unambiguous events (Figure \ref{fig:DetailedStages}), identifiable regardless of embryo orientation, were selected for refinement and further analysis (pole bud appears, membrane reaches yolk, pole cell invagination, amnioproctodeal invagination, amnioserosa exposed, clypeolabrum retracts, heart-shaped midgut, and trachea fill) (Figure \ref{fig:AutomatedEventPrediction}B,C). Using a Python-scripted graphical user interface, each of the eight events in every movie was manually examined and the algorithm prediction adjusted when necessary. Timing of hatching was excluded from these nine primary events because it was highly variable, likely due to the assay conditions following dechorionation, and suitable only as an indication of successful development, not as a reliable and reproducible time point. The ``membrane reaches yolk stage'' was used throughout as a zero point due to the precision with which the stage could be identified in all species and from all orientations.

Links to representative time-lapse videos are provided in Table \ref{tab:VideoList}.

\subsection*{\textit{D. melanogaster} embryogenesis scales uniformly with temperature}

\begin{figure}[!ht]
\begin{center}
\includegraphics[width=5.5in]{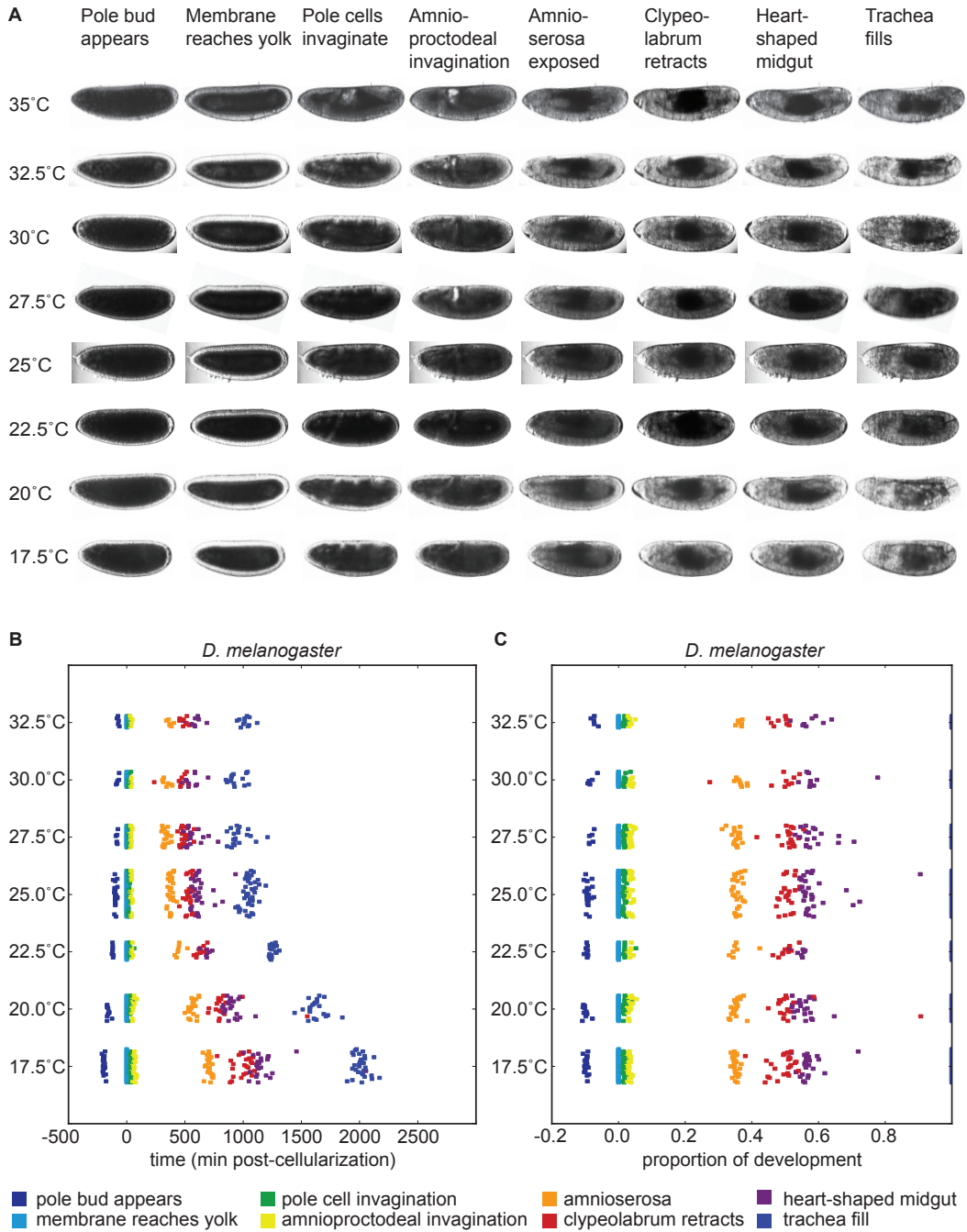}
\end{center}
\caption{\textbf{Developmental time of \textit{D. melanogaster} varies with temperature} (A) Images of developing \textit{D. melanogaster} embryos at each temperature are shown for a selection of stages to highlight the overall similarity of development. (B) The time individual animals reached the various time-points are shown, with each event being a different color. Time 0 is defined as the end of cellularization, when the membrane invagination reaches the yolk. Between 17.5\degree C and 27.5\degree C the total time of embryogenesis, $t_{dev}$ measured as the mean time between cellularization and trachea fill, has a logarithmic relationship to temperature described by $t_{dev} = 4.02 e^{ 37.31 / T}$ where T is temperature in \degree C ($R^2 = 0.963$). (C) The developmental rate in \textit{D. melanogaster} changes uniformly with temperature, not preferentially affecting any stage. Timing here is normalized between the end of cellularization and the filling of the trachea. }
\label{fig:DmelAcrossT}
\end{figure}

As expected, the total time of embryogenesis of \textit{D. melanogaster} had a very strong dependence on temperature (Figure \ref{fig:DmelAcrossT}, \url{http://www.youtube.com/watch?v=-yrs4DcFFF0} or ``\textit{D. melanogaster} at 7 temperatures'' in DOI:10.5061/dryad.s0p50). From 17.5\degree C to 27.5\degree C, there was a two-fold acceleration in developmental rate, matching the previously observed doubling of total lifespan with a 10\degree C change in temperature \cite{Loeb1916}. The velocity of embryogenesis at 30\degree C is roughly the same as at 27.5\degree C, and is appreciably slower at 32.5\degree C, likely due to heat stress. At 35\degree C, successful development becomes extremely rare.

To examine how these temperature-induced shifts in the total time of embryogenesis were reflected in the relative timing of individual events, we rescaled the time series data for each embryo so that the time from our most reliable early landmark (the end of cellularization) to our most reliable late landmark (trachea filling) was identical, and examined where each of the remaining landmarks fell (Figure \ref{fig:DmelAcrossT}C). We were surprised to find that \textit{D. melanogaster} exhibited no major changes in its proportional developmental time under any of the non-stressful temperature conditions tested. Therefore, at least as far as most visually evident morphological features go, embryogenesis scales uniformly across a two-fold range of total time. 

When the embryos were under heat stress ($>$30\degree C), we observed a very slight contraction in the proportion of time between early development (pole bud appears) to the end of cellularization (membrane reaches yolk), and a slight contraction between the end of cellularization and mid-germ band retraction (amnioserosa exposure). 
		
\subsection*{Embryogenesis scales uniformly across species despite significant differences in temperature dependent developmental rate}

\begin{figure}[!ht]
\begin{center}
\includegraphics[width=5.5in]{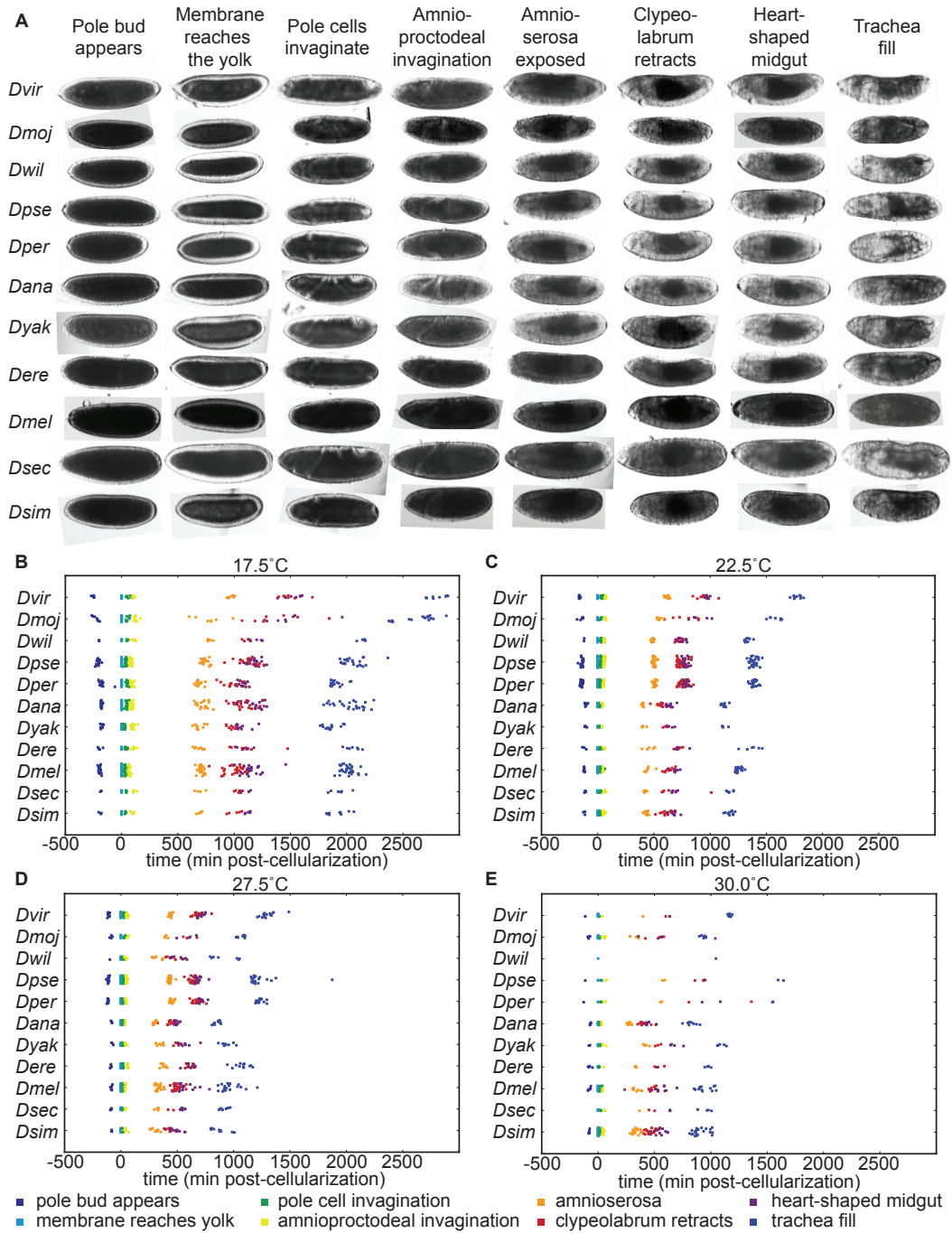}
\end{center}
\caption{\textbf{\textit{Drosophila} species develop at different rates and respond to temperature in distinct ways} (A) Images of developing embryos of each species are shown to scale. All species go through the same stages in the same order at all viable temperatures. (B) At 17.5\degree C all species show uniformly long developmental times, with \textit{D. virilis} and \textit{D. mojavensis} being significantly longer than other species. (C) At 22.5\degree C and (D) 27.5\degree C there is considerably more variation between species. While developmental times decrease with increasing temperature across all species, the effect is muted in the alpine species. (E) At 30\degree C, developmental rate has stopped accelerating and the alpine species are seeing considerable slow-down in development time. }
\label{fig:DvarAcrossT}
\end{figure}

\begin{figure}[!ht]
\begin{center}
\includegraphics[width=5.5in]{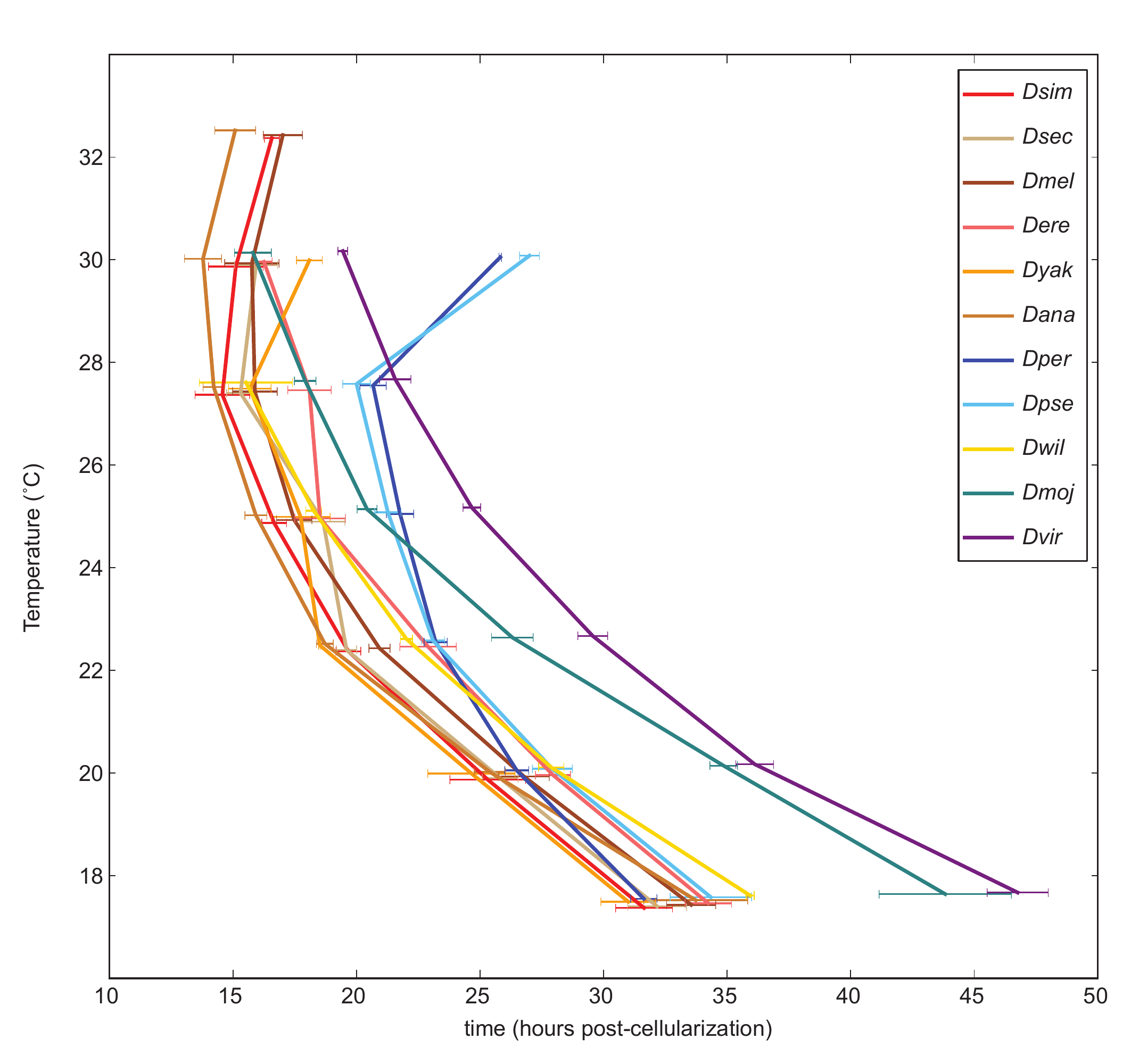}
\end{center}
\caption{\textbf{Temperature dependent developmental rates are climate specific} The time between the end of cellularization and trachea fill are shown for all species at a range of temperatures. The climatic groups -- tropical (warm colors), alpine (blues), temperate (purple), and sub-tropical (green) -- clearly stand out from one another to form four general trends.}
\label{fig:TacrossSp}
\end{figure}

In each of the ten additional \textit{Drosophila} species we examined we observed all of the 36 developmental landmarks we identified in \textit{D. melanogaster} in the same temporal order (Figure \ref{fig:DvarAcrossT}A). However, there was marked interspecies variation in both the total time of embryogenesis at a given temperature (Figure \ref{fig:DvarAcrossT}B-E, Table \ref{tab:VideoList}) and the way embryogenesis time varied with temperature (Figure \ref{fig:TacrossSp}). 

\begin{figure}[!ht]
\begin{center}
\includegraphics[width=5.5in]{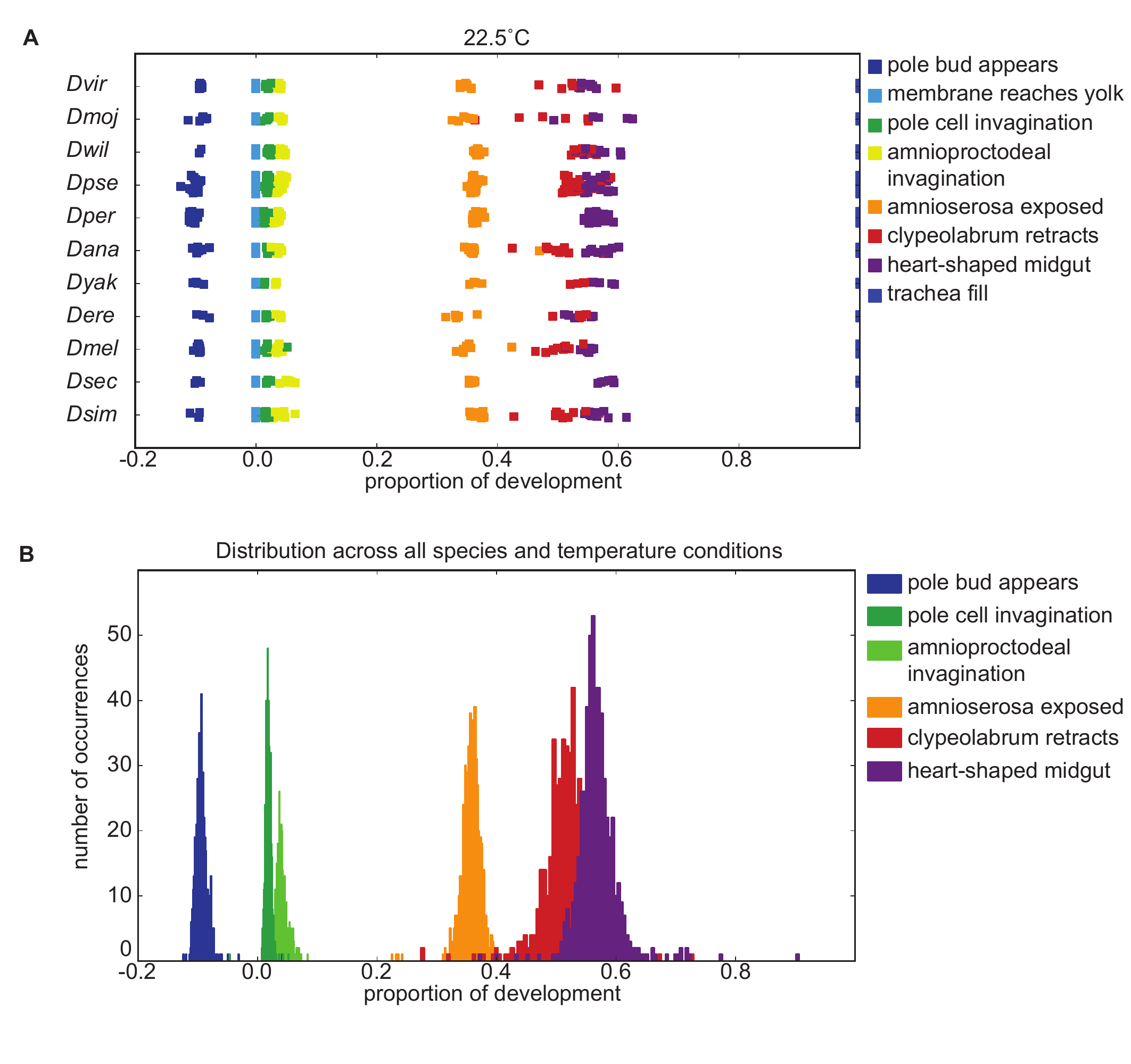}
\end{center}
\caption{\textbf{Proportionality of developmental stages is not affected by non-heat-stress temperatures} (A) Across species, development maintains the same proportionality. \textit{D. pseudoobscura} stands out as not being co-linear at higher temperatures. Instead, the later part of its development is slowed and takes up a disproportionally long time. (B) Plotting proportionality across all species and all temperatures reveals the approximately normally distributed proportionality of all morphological stages. }
\label{fig:Proportionality}
\end{figure}

When we examined the 10 remaining species, we found not only that the relative timing of events was constant across temperature within a species, as observed in \textit{D. melanogaster}, but that landmarks occurred at the same relative time between species at all non-stressful temperatures (Figures \ref{fig:Proportionality}, Table \ref{tab:EventTiming}). 

\subsection*{Developmental time is exponentially related to $\alpha/T$}

\begin{figure}[h]
\includegraphics[scale=0.75]{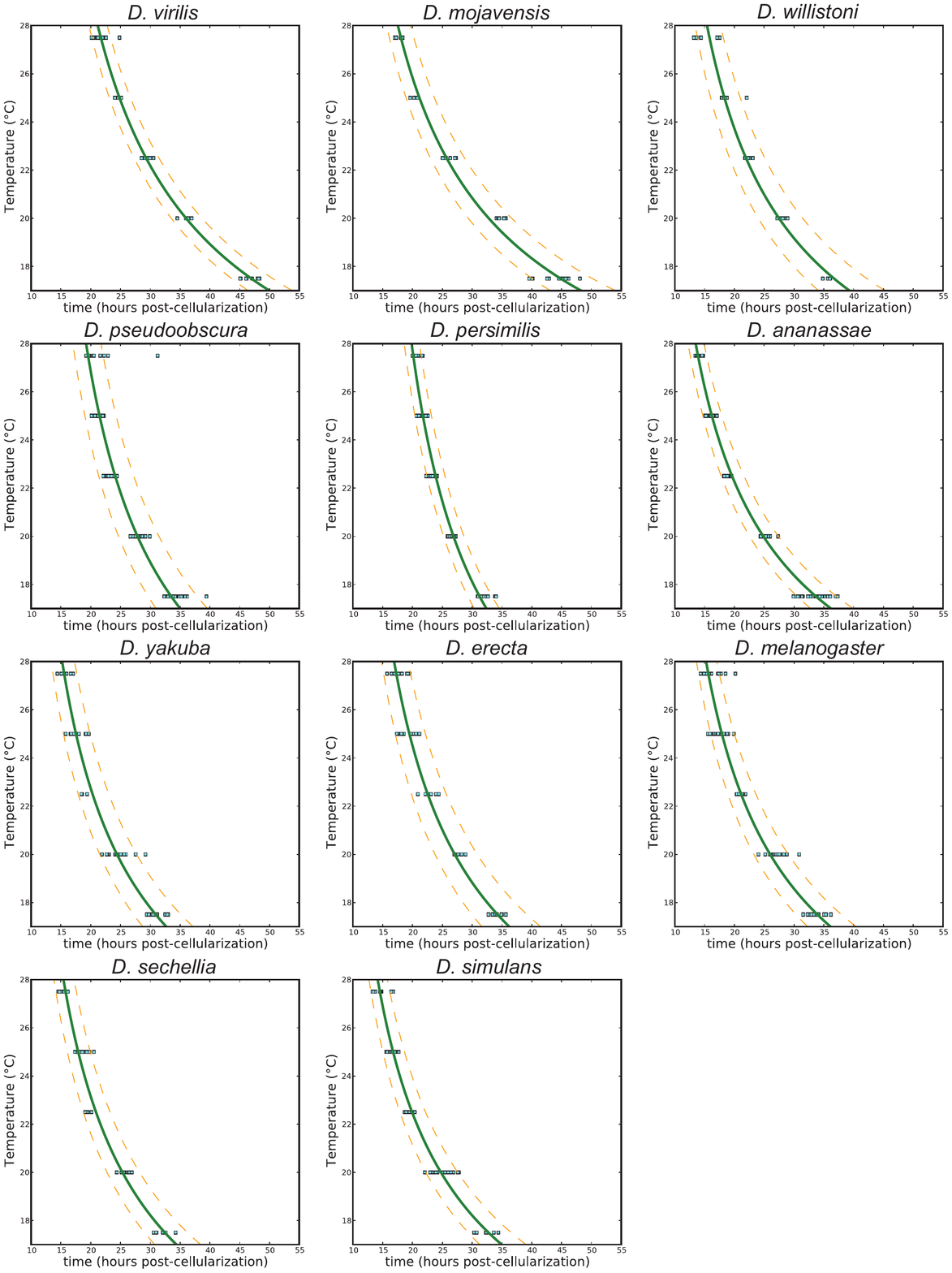}
\caption{\textbf{Prediction of future observations of development at different temperatures} The behavior of developing embryos can be predicted. The mean line (green) generated from least-squares curve-fitting (Table \ref{tab:SpeciesTiming}) and the 95\% confidence prediction interval for future observations (dashed orange line) are shown for each species.}
\label{fig:tPredictionsFromT}
\end{figure}

Between 17.5\degree C and 27.5\degree C the total developmental time for all species can be approximated relatively accurately by an exponential regression ($R^2 > 0.9$). For all species we find that temperature \textit{T} can be related to developmental time $t_{dev}$, agreeing with a long history of temperature-dependent rate modeling \cite{Arrhenius:1915}: \[t_{dev} \approx e^\frac{\alpha}{T}\] and developmental rate \textit{v}: \[ln(v) \approx -\frac{\alpha}{T}\] The parameters of these relations for each species, which includes two independent coefficients, are included in Table \ref{tab:SpeciesTiming}. Also included in Table \ref{tab:SpeciesTiming} is the $Q_{10}$, an empirical description of biological rate change from a 10\degree C temperature change, for the 17.5\degree C to 27.5\degree C interval. At higher temperatures, heat stress appears to counter the logarithmic trend and lengthens developmental time. Since the temperature responses are highly reproducible, the developmental time for each species can be modeled and predictions made for future experiments (Figure \ref{fig:tPredictionsFromT}).

\subsection*{Effect of temperature on developmental rate is coupled to climatic origin }

Seven of the eleven species we examined were of tropical origin, with only two alpine, one subtropical and one temperature species. At mid-range temperatures (22.5\degree C - 27.5\degree C), the tropical species developed the fastest, followed by the subtropical \textit{D. mojavensis}, the alpine \textit{D. pseudoobscura} and \textit{D. persimilis}, and the temperate \textit{D. virilis} (Figure \ref{fig:TacrossSp}), in accord with \cite{Markow:2009}. 

\begin{figure}[h]
\includegraphics[scale=0.65]{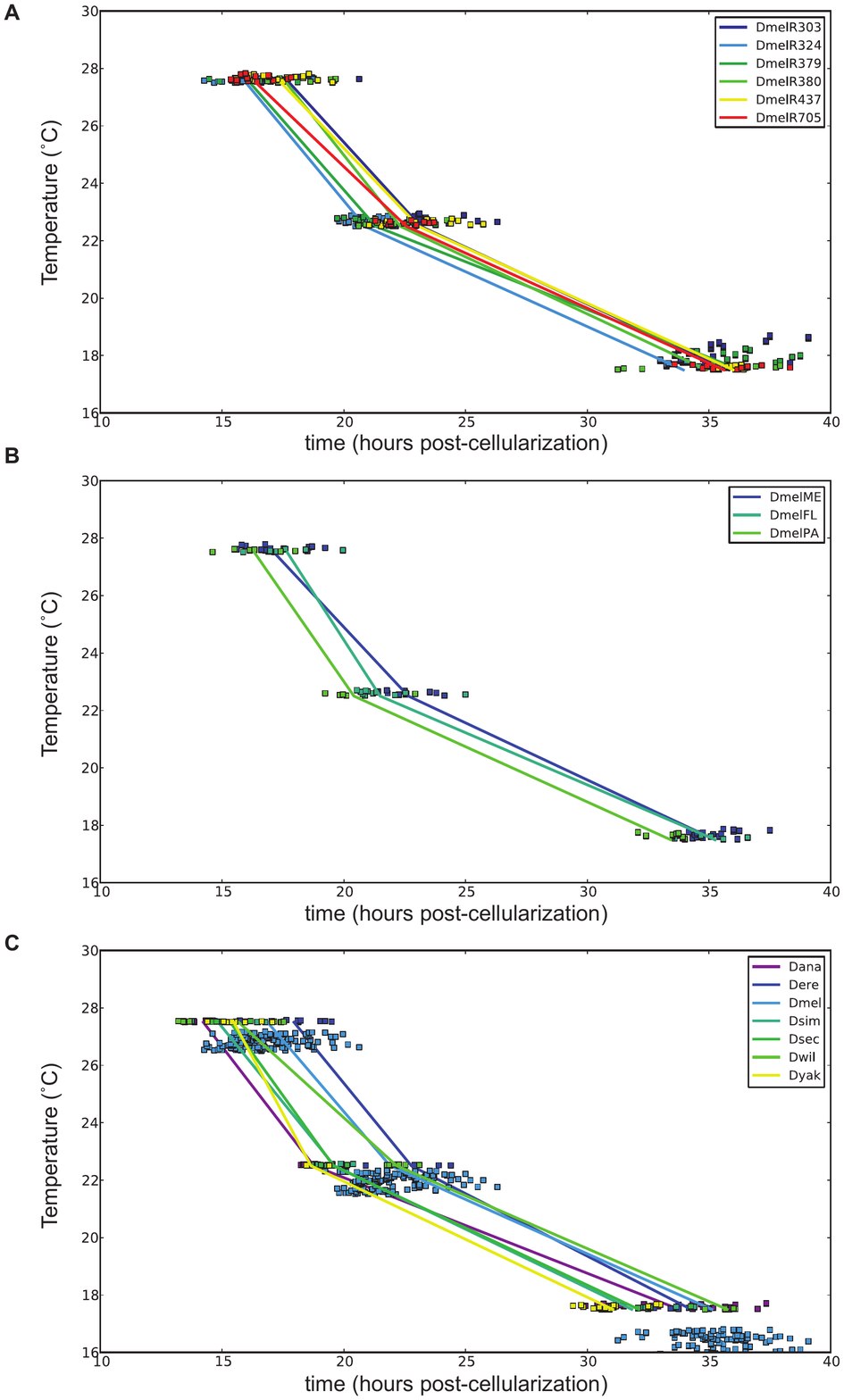}
\caption{\textbf{Different \textit{D. melanogaster} wild isolate strains exhibit a limited range of temperature responses} (A) Lines (R303, R324, R379, R380, R437, and R705) collected near Raleigh, North Carolina \cite{Mackay:2012fd} exhibit a range of temperature responses. (B) Clinal lines from Florida (DmelFL), Pennsylvania (DmelPA), and Maine (DmelME) \cite{Fabian:2012} exhibit a range of responses similar to those of the Raleigh lines. Despite their clinal distribution, no trends are seen, with flies from Florida and Maine being virtually indistinguishable. This is possibly due to their relatively recent introduction across the cline. (C) Despite the differences between the \textit{D. melanogaster} lines above, they all (seen here grouped together as light blue points) lie within the response range seen for the \textit{melanogaster} species subgroup, mainly falling between the responses of \textit{D. melanogaster} Ore-R and \textit{D. erecta}. Like Ore-R, their growth is significantly slower than \textit{D. yakuba}, \textit{D. ananassae}, \textit{D. simulans}, and \textit{D. sechellia}, but obeys the same general trend.}
\label{fig:StrainDifferences}
\end{figure}

Some tropical species have expanded into temperature zones and a variety of wild strains have been collected from a variety of climates. We examined nine additional strains of \textit{D. melanogaster} collected along the eastern United States \cite{Mackay:2012fd, Fabian:2012}. Though collected along a tropical to temperate cline and there was some variation between strains, no trends were seen (Figure \ref{fig:StrainDifferences}A,B).

\begin{figure}[h]
\includegraphics[scale=0.7]{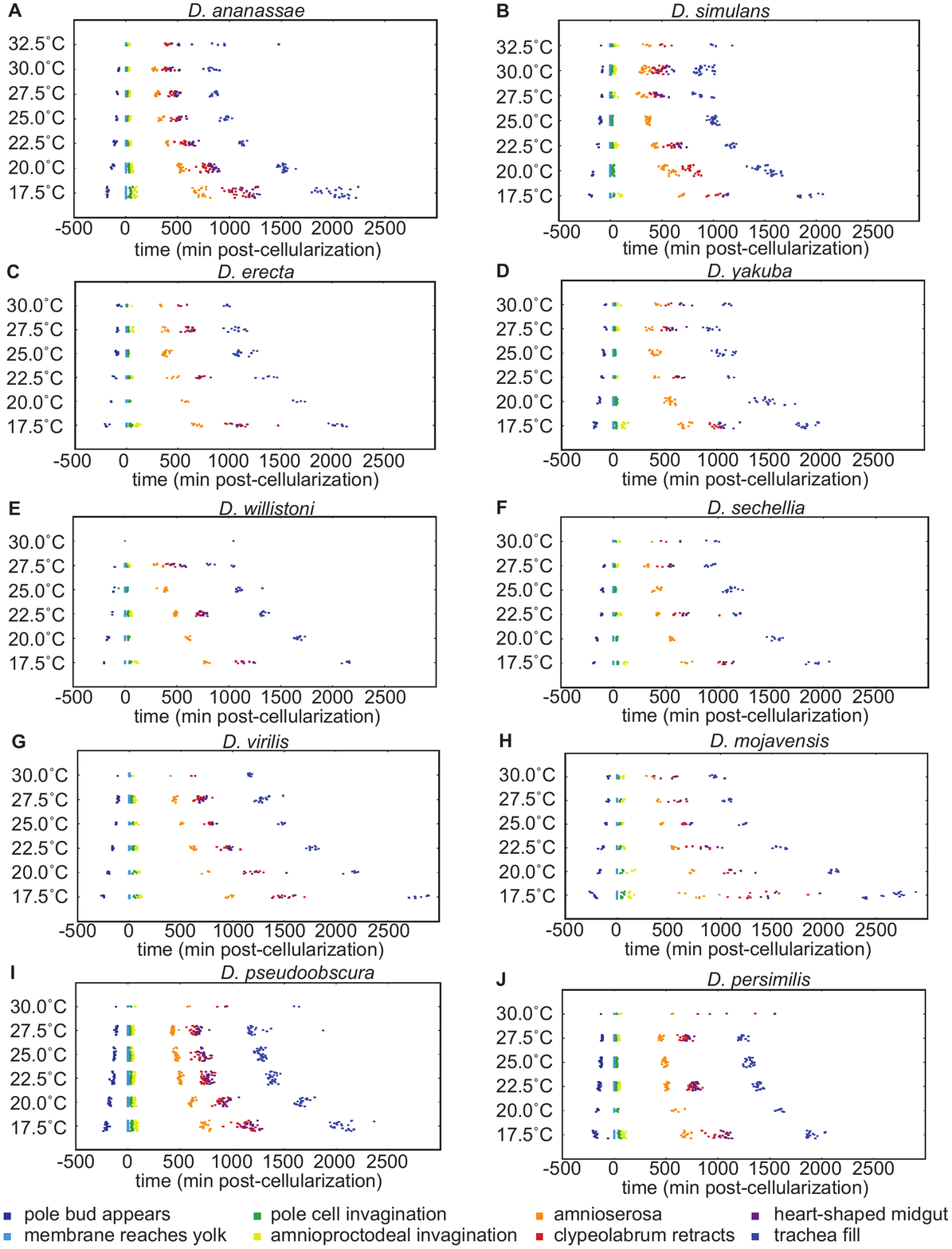}
\caption{\textbf{Ten species of \textit{Drosophila} exhibit dynamic response to temperature changes} (A-F) There is some variation species to species, but all tropical \textit{Drosophila} exhibit a similar temperature response-curve to \textit{D. ananassae}. (G) Temperate \textit{D. virilis} also has a steep response, though intermediate to the previous two groups. (H) Sub-tropical \textit{D. mojavensis} has a steeper temperature response, though a similar high temperature developmental time. (I,J) Alpine \textit{D. pseudoobscura} and \textit{D. persimilis} have a cold response like the tropical species, but longer developmental times at warmer temperatures.}
\label{fig:DspAcrossT}
\end{figure}

The tropical species all showed highly similar responses to temperature, even though they originate from different continents (Africa, Asia and South America) and are not closely related (five of the species are in the melanogaster subgroup, but \textit{D. ananassae} and \textit{D. willistoni} are highly diverged from both \textit{D. melanogaster} and each other). Though they possess similar temperature-responses, these species possess significantly different and independent temperature response curves ($p < 0.05$) and the differences are large enough to be relevant for precise developmental experiments. These cross-species differences tend to be, but are not necessarily, larger than those seen between \textit{D. melanogaster} strains (Figure \ref{fig:StrainDifferences}C). The embryogenesis rate for these species increases rapidly with temperature ($Q_{10} \sim 2.2$) before slowing down at and above 30\degree C (Figure \ref{fig:DspAcrossT}A-F, \url{http://www.youtube.com/watch?v=vy6L4fmWkso} or ``\textit{D. ananassae} at 7 temperatures'' in DOI:10.5061/dryad.s0p50). The two closely related alpine species (\textit{D. pseudoobscura} and \textit{D. persimilis}) match the embryogenesis rate of the tropical species at 17.5\degree C, but accelerate far less rapidly with increasing temperature ($Q_{10} \sim 1.6$), especially at 25\degree C and above (Figure \ref{fig:DspAcrossT}I,J, \url{http://www.youtube.com/watch?v=sYi-FUXpv4Q} or ``\textit{D. pseudoobscura} at 6 temperatures'' in DOI:10.5061/dryad.s0p50). These species also show a sharp increase in embryogenesis rate and low viability above 27.5\degree C, consistent with their cooler habitat.  The subtropical \textit{D. mojavensis} (Figure \ref{fig:DspAcrossT}H, \url{http://www.youtube.com/watch?v=XWMs4oUx_mU} or ``\textit{D. mojavensis}  at 6 temperatures" in DOI:10.5061/dryad.s0p50) and temperate \textit {D. virilis} (Figure \ref{fig:DspAcrossT}G, \url{http://www.youtube.com/watch?v=eyr4ckDb0kM} or ``\textit{D. virilis} at 6 temperatures'' in DOI:10.5061/dryad.s0p50) both develop very slowly at low temperature, but accelerate rapidly as temperature increases ($Q_{10}$ of $\sim 2.5$ and $\sim 2.2$ respectively). \textit {D. virilis} remains the slowest species up to 30\degree C, while \textit{D. mojavensis} is as fast as the tropical species at high temperatures. These species are both members of the \textit{virilis-repleta} radiation and it remains to be seen if this growth response is characteristic of the group as a whole, independent of climate.

\subsection*{Effects of heat stress}

\begin{figure}[h]
\includegraphics[scale=0.65]{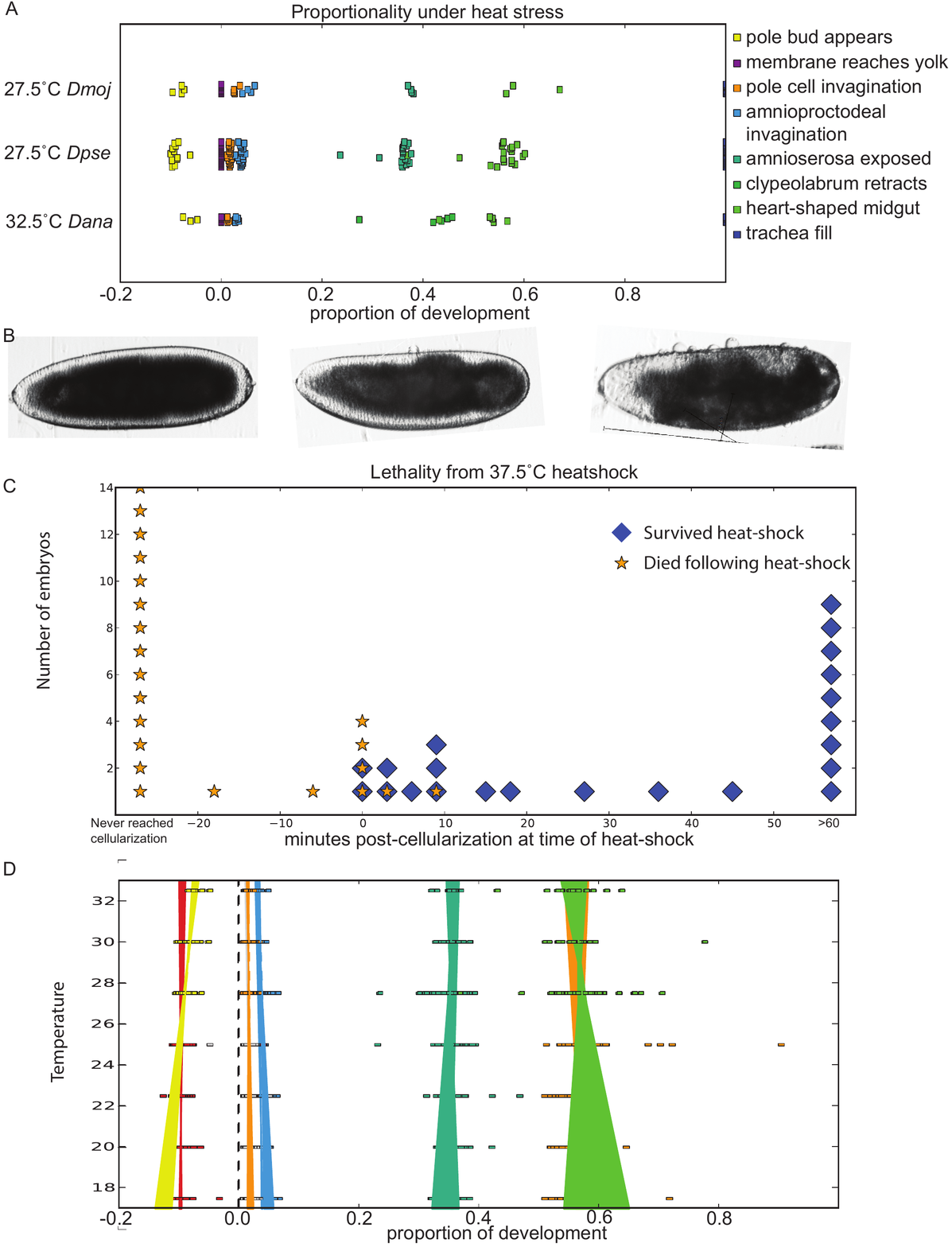}
\caption{\textbf{Heat-stress affects syncytial developmental proportionality and morphology} (A) At heat-stress temperatures, the proportionality of developmental stages is affected in some, but not all, embryos. (B) Heat stress in \textit{D. melanogaster} at 32.5\degree C affects morphology during yolk contraction and gastrulation. Embryos may exhibit asynchronous yolk-contraction (first image), uneven nuclear distribution during cellularization (second image), or disrupted morphology during gastrulation (third image). (C) Heat shock at 37.5\degree C for 30 minutes reveals embryos sensitivity prior to the completion of cellularization. Most animals that had completed cellularization survived heat-shock and continued to develop properly (blue diamonds), while no animals that had not completed cellularization prior to heat-shock survived. All embryos that died (orange stars) exhibited severe morphological disruptions. (D) Linear regression of stages across different temperatures reveals that, despite significant variance in later stages (shown in colored bars), only the pre-cellularization time point is affected by heat-stress enough to exhibit a significantly different slope between higher temperatures (27.5\degree C and above, yellow bar) and lower temperatures (25\degree C and below, red bar). }
\label{fig:TrendsInProportionality}
\end{figure}

Under heat-stress, the proportionality of development is disrupted in some embryos (Figure \ref{fig:TrendsInProportionality}A). The effect is not uniform, as some embryos developed proportionally under heat-stress and others exhibited significant aberrations, largely focused in post-germband shortening stages. This can be most clearly seen in individuals of \textit{D. ananassae}, \textit{D. mojavensis}, \textit{D. persimilis}, and \textit{D. pseudoobscura}.  We did not identify any particular stage as causing this delay, but rather it appears to reflect a uniform slowing of development. 

Early heat shock significantly disrupts development enough to noticeably affect morphology in yolk contraction, cellularization, and gastrulation (Figure \ref{fig:TrendsInProportionality}B). Syncytial animals are the most sensitive to heat-shock (Figure \ref{fig:TrendsInProportionality}C). In \textit{D. melanogaster} and several other species we observed a slight contraction of proportional developmental time between early development (pole bud appears) and the end of cellularization (membrane reaches yolk) under heat-stress ($>$ 30\degree C, Figure \ref{fig:TrendsInProportionality}D). While all later stages following cellularization maintain their proportionality even at very high temperatures, the pre-cellularization stages take proportionally less and less time. This indicates that at higher temperatures, some pre-cellularization kinetics scale independently of later stages, possibly leading to mortality as the temperature becomes more extreme.

\section*{Discussion}
We have addressed the lack of good data on the progress of embryogenesis in different species and at different temperatures with a carefully collected and annoted series of time-lapse movies in 11 species at seven temperatures that span most of the viable range for \textit{Drosophila} species. From a practical standpoint, the predictable response of each species to temperature, and the uniform scaling of events between species and temperature, provides a relatively simple answer to the question that motivated this study - to determine how to obtain matched samples for genomic studies: simply choose the range of stages to collect in one strain or species, and scale the collection and aging times appropriately. The fact that development scales uniformly over non-extreme temperatures would seem to give some leeway in the choice of temperature, so long as heat-stress is avoided, though it remains unclear how molecular processes are affected by temperature. 

\subsection*{Uniform scaling and the timing of embryogenesis}

In carrying out this survey, we were surprised to find that the relative timing of landmark events in \textit{Drosophila} embryogenesis is constant across greater than three-fold changes in total time, spanning 15\degree C and over 100 million years of independent evolution. And the fact that the same holds true for 34 developmental landmarks at two temperatures in the zebrafish \textit{Danio rerio} \cite{Kimmel:1995sed}, (the only other species for which we were able to locate similar data), suggests that this phenomenon may have some generality. But why is this so? 

\textit{Drosophila} development involves a diverse set of cellular processes including proliferation, growth, apoptosis, migration, polarization, differentiation, and tissue formation. One might expect (we certainly did) these different processes to scale independently with temperature, much as different chemical reactions do, and as a result, different stages of embryogenesis or parts of the developing embryo would scale differentially with temperature. But this is not the case. The simplest explanation for this observation is that a single shared mechanism controls timing across embryogenesis throughout the genus \textit{Drosophila}. But what could such a mechanism be? One possibility is that there is an actual clock \textemdash  some molecule or set of molecules whose abundance or activity progresses in a clocklike manner across embryogenesis and is read out to trigger the myriad different processes that occur in the transition from a fertilized egg to a larvae. However there is no direct evidence that such a clock exists (although we note that there is a pulse of ecdysone during embryogenesis with possible morphological functions \cite{Riddiford:1993, Chavez:2000ecd}). A more likely explanation is that there is a common rate limiting process throughout embryogenesis. Our data are largely silent on what this could be, but we know from other experiments that it is cell, or at least locally, autonomous \cite{Lucchetta:2005ep, Niemuth:1995, Girdler:2013} and would have to limit processes like migration that do not require cell division (we also note that cell division has been excluded as a possibility in zebrafish \cite{Girdler:2013}). However, energy production, yolk utilization, transcription or protein synthesis are reasonable possibilities. 

Although there are very few comparisons of the relative timing of events during development, it has long been noted that various measurements of developmental timing scale exponentially with $\alpha/T$ \cite{Lillie:1897, Arrhenius:1915, Loeb1916, Powsner:1935, Davidson:1942}, but no good explanation for this phenomenon has been uncovered. Perhaps development is more generally limited by something that scales exponentially with $\alpha/T$, like metabolic rate, which, we note, has been implicated numerous times in lifespan, which is, in some ways, a measure of developmental timing. 

Gillooly and co-workers, noting the there was a relationship between metabolic rate, temperature and animal size, have proposed a model that incorporates mass into the Arrhenius equation to explain the relationship between these factors in species from across the tree of life \cite{Gillooly:2001, Gillooly:2002na}. We, however, do not find that mass can explain the differences in temperature-dependence between species.  Even closely-related species, with nearly 2-fold differences in their mass (e.g. \textit{D. melanogaster}, \textit{D. simulans}, \textit{D. sechellia}, \textit{D. yakuba}, and \textit{D. erecta}), have significant divergence in their proportionality coefficients that do not converge at all when correcting for differences in mass through the one quarter power scaling proposed by Gillooly, et al. This suggests that some other factor is responsible for the differences, as has been argued by other groups \cite{Clarke:2004sa, Clarke:2004b, Markow:2009}. The relationship between climate and temperature response raises the possibility that whatever this factor is has been subject to selection to tune the temperature response to each species' climate. However, without additional data this is purely a hypothesis. 

Although a common rate-limiting step is simplest explanation for uniform scaling, it is certainly not the only one. It is possible that different rate limiting steps or other processes control developmental velocity at different times or in different parts of the embryo, and that they scale identically with temperature either coincidentally, or as the result of selection (it is important to remember that, as per Arrhenius, one does not expect different reactions to scale identically with temperature). If this is the result of selection, what is the selection pressure? Evolutionary developmental biologists, perhaps most notably Stephen J. Gould, have long written about how changes in either the absolute or relative timing of different events during development have had significant effects on morphology throughout animal evolution \cite{McNamara:1982, Jones:1999, Felix:1999, Patel:1994pa}. Perhaps this is also true for fly embryogenesis, but that any such changes in morphology are selectively disadvantageous and have been strongly selected against. It is also likely that many developing fly embryos experience significant changes in temperature while developing, so there may be strong selection to maintain uniform development across temperature to ensure normal progression while the temperature is changing.

Finally, we note that there are limits to this uniformity. At extreme temperatures, especially high ones, things no longer scale uniformly, likely reflecting the differential negative effects of high temperature at different stages of embryogenesis as well as the differential ability of the embryo to compensate for them. There are also clearly checkpoints in place that, while not triggered during normal embryogenesis, are important in extreme or unusual circumstances. Most strikingly, when Lucchetta et al. and Niemuth et al. examined embryos developing in chambers that allowed for independent temperature control of the anterior and posterior portions of the embryo, the two parts of the embryo developed at different velocities for much of embryogenesis \cite{Lucchetta:2005ep, Niemuth:1995}. They found that embryos are robust to asynchrony in timing across the embryo, though there are critical periods that, once passed, do not permit re-synchronization of development \cite{Lucchetta:2005ep}, hinting at some specific checkpoints or feedback.

\subsection*{Climate and the rate of embryogenesis}

The clustering of developmental timing and its temperature response with climate \textemdash especially amongst tropical species from different continents and parts of the \textit{Drosophila} tree \textemdash suggests that this is an adaptive, or in some cases permissive, phenotype, although with only 11 species and poor coverage of non-tropical species this has to remain highly speculative. There are necessarily additional components to the temperature response, as significant variation exists within the tropical species and between \textit{D. melanogaster} strains. The \textit{virilis-repleta} radiation, which includes both \textit{D. virilis} and \textit{D. mojavensis} may have a climate-independent adaptation that leads to slowed development at cooler temperatures, a feature that is hard to rationalize. The poor response of the alpine \textit{D. pseudoobscura} and \textit{D. persimilis} to high temperature is consistent with their cool climate. Nevertheless, little is known about when and where most of these species lay their eggs and their natural microclimates.

The clustering of developmental responses in species by their native climates rather than their climates of collection suggests that if climate adaptation is a contributing factor, the response arises slowly or rarely. The tested strains of \textit{D. melanogaster} were collected in temperate, subtropical, and tropical climates and the \textit{D. simulans} strain was collected in a sub-tropical climate. Nevertheless, both species performed qualitatively like other tropical species and unlike native species collected nearby. This suggests that temperature responses are neither rapidly evolving (with \textit{D. melanogaster} being present in the temperate United States for over 130 years \cite{Keller:2007wo}) nor primed for change in tropical species. 

\section*{Materials and Methods}
\subsection*{Rearing of \textit{Drosophila}}
\begin{figure}[h]
\includegraphics[scale=0.70]{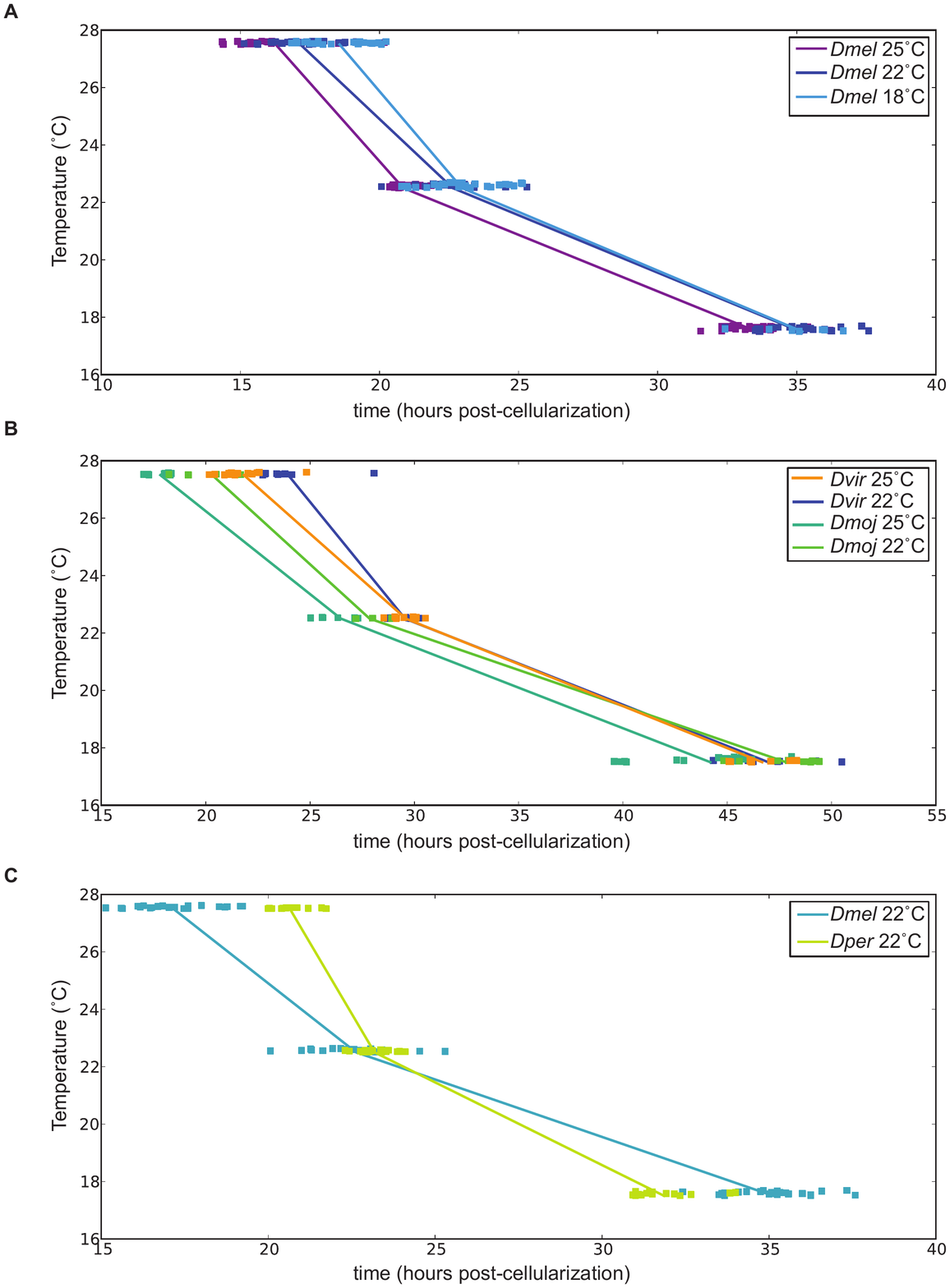}
\caption{\textbf{Temperature conditioning of adult flies leads to some heat tolerance} (A) \textit{D. melanogaster} raised for many generations at 25 \degree C, 22 \degree C, and 18 \degree C produce embryos that show similar temperature responses, though there is some accelerated growth when acclimatized to higher temperatures. There is no indication of severe heat shock as embryos are moved from the acclimatized temperature to the experimental temperature. (B) \textit{D. mojavensis} and \textit{D. virilis} exhibit a similar trend of only minor differences between strains acclimatized at 25 \degree C and 22 \degree C. (C) The difference between \textit{D. melanogaster} raised at 22 \degree C and \textit{D. persimilis} also raised at 22 \degree C remains significant, indicating that the heat-stress response of \textit{D. persimilis} is not due simply to its being raised at 22 \degree C.}
\label{fig:TemperatureConditioning}
\end{figure}

\textit{Drosophila} strains were reared and maintained on standard fly media at 25\degree C, except for \textit{D. persimilis} and \textit{D. pseudoobscura} which were reared and maintained at 22\degree C. \textit{D. melanogaster} lines were raised at 18\degree C and 22\degree C for several years and their temperature response profiles were observed, verifying that transferring embryos from the ambient growth temperature for a line to the experimental temperature did not lead to heat-shock responses and had relatively little impact on the temperature response (Figure \ref{fig:TemperatureConditioning}A,B). Egg-lays were performed in medium cages on 10 cm molasses plates for 1 hour at 25\degree C after pre-clearing for all species except \textit{D. persimilis}, which layed at 22\degree C. Comparisons to \textit{D. melanogaster} raised and laying at 22\degree C confirmed that growth at lower temperatures does not account for all of the differences between the tropical and alpine species (Figure \ref{fig:TemperatureConditioning}C).To encourage egg-lay, cornmeal food media was added to plates for \textit{D. sechellia} and pickled cactus was added to plates for \textit{D. mojavensis}. Embryos were collected and dechorionated with fresh 50\% bleach solution (3\% hypochlorite final) for 45 to 90 seconds (based on the species) in preparation for imaging. Dechorionation timing was selected as the time it took for 90\% of the eggs to be successfully dechorionated. This prevented excess bleaching, as many species, such as \textit{D. mojavensis}, are more sensitive than \textit{D. melanogaster}. Strains used were \textit{D. melanogaster}, OreR, DGRP R303, DGRP R324, DGRP R379, DGRP R380, DGRP R437, DGRP R705, Schmidt Ln6-3, Schmidt 12BME10-24, and Schmidt 13FSP11-5; \textit{D. pseudoobscura}, 14011-0121.94, MV2-25; \textit{D. virilis}, 15010-1051.87, McAllister V46; \textit{D. yakuba}, 14021-0261.01, Begun Tai18E2; \textit{D. persimilis}, 14011-0111.49,(Machado) MSH3; \textit{D. simulans}, 14021-0251.195, (Begun) simw501; \textit{D. erecta}, 14021-0224.01, (TSC); \textit{D. mojavensis wrigleyi}, 15081-1352.22, (Reed) CI 12 IB-4 g8;  \textit{D. sechellia}, 14021-0248.25, (Jones) Robertson 3C; \textit{D. willistoni}, 14030-0811.24, Powell Gd-H4-1; \textit{D. ananassae}, 14024-0371.13, Matsuda (AABBg1).

\subsection*{Time-lapse Imaging}
Embryos were placed on oxygen-permeable film (lumox, Greiner Bio-one), affixed with dried heptane glue  and then covered with Halocarbon 700 oil (Sigma) \cite{Technau1986}. The lumox film was suspended on a copper plate that was temperature-regulated with two peltier plates controlled by an H-bridge temperature controller (McShane Inc., 5R7-570) with a thermistor feedback, accurate to \plusminus 0.1\degree C. Time-lapse imaging with bright field transmitted light was performed on a Leica M205 FA dissecting microscope with a Leica DFC310 FX camera using the Leica Advanced Imaging Software (LAS AF) platform. Greyscale images were saved from pre-cellularization to hatch. Images were saved every one to five minutes, depending on the temperature. A humidifier was used to mitigate fluctuations in ambient humidity, though fluctuations did not affect developmental rate. Due to fluctuations in ambient temperature and humidity, the focal plane through the halocarbon oil varied significantly. Therefore, z-stacks were generated for each time-lapse and the most in-focus plane at each time was computationally determined for each image using an algorithm (implemented in Matlab) through image autocorrelation \cite{Santos1997, Vollath1988}. Time-lapse videos available from Dryad Digital Repository: doi:10.5061/dryad.s0p50

\subsection*{Event estimation}
A subset of time-lapses in \textit{D. melanogaster} were analyzed to obtain a series of representative images for each of the 36 morphological events, selected as all events defined by \cite{Campos-Ortega1985, Bownes1975} that were reproducibly identifiable under our conditions, described. These images were sorted based on embryo orientation and superimposed to generate composite reference images. Images from each time-lapse to be analyzed were manually screened to determine the time when the membrane reaches the yolk, the time of trachea filling, and the orientation of the embryo (Figure \ref{fig:DetailedStages}. This information was fed into a Matlab script, along with the time-lapse images and the set of 34 composite reference images, to estimate the time of 34 morphological events during embryogenesis via image correlation. The same \textit{D. melanogaster} reference images were used for all species for consistency. A correlation score was generated for each frame of the time-lapse. The running score was then smoothed (Savitzky-Golay smoothing filter) and the expected time window was analyzed for local maxima. The error in event calling for the computer is very large (greater than what we see for the overall spread across individuals of a single species at a given temperature), necessitating manual verification or correction of events. Many of these errors are due to aberrations in the image that confuse the computer but would not confuse a person. This results in a few bad images having a very negative effect of the overall accuracy of the computer analysis, but permits a significant improvement with just a little user input. The error in manual calls is very small compared to the variation between individuals. Computer-aided estimates were individually verified or corrected using a python GUI for all included data.

\subsection*{Statistical analysis}
Statistical significance of event timing was determined by t-test with Bonferonni multiple testing corrections. Median correction to remove outliers was used in determining the mean and standard deviation of each developmental event. Least-squares fitting was used to determine the linear approximation of log-corrected developmental time for each species. Python and Matlab scripts used in the data analysis are available at github.com/sgkuntz/TimeLapseCode.git.

\section*{Acknowledgments}
We obtained flies from the Bloomington and UCSD stock centers. We thank Paul Schmidt for the clinal fly lines. We would also like to thank T. Kaplan and P. Combs for their advice and assistance on data analysis and programming, C. Bergman for comments, Ng Wei Tian for his work on event verification, and BCK and EDKK for their support.

\bibliography{WholeLibrary}

\section*{Figures}

\clearpage

\section*{Tables}

\begin{table}[!ht]
\caption{\bf{\textit{Drosophila} species and strains}}
\begin{tabular}{| l | l | l | l |}

	\hline
	Species & Stock number & Strain & Collection site \\
	\hline
	\textit{D. melanogaster} & & OreR & Oregon, USA \\
	\textit{D. pseudoobscura} & 14011-0121.94 & MV2-25 & Mesa Verde, Colorado, USA \\
	\textit{D. virilis} & 15010-1051.87 & McAllister V46 & \textit{unknown, possibly Asia} \\
	\textit{D. yakuba} & 14021-0261.01 & Begun Tai18E2 & Liberia \\
	\textit{D. persimilis} & 14011-0111.49 & Machado MSH3 & Mt. St. Helena, California, USA \\
	\textit{D. simulans} & 14021-0251.195 & Begun simw501 & Mexico City, Mexico \\
	\textit{D. erecta} & 14021-0224.01 & (TSC) & \textit{unknown, probably Africa} \\
	\textit{D. mojavensis wrigleyi} & 15081-1352.22 & Reed CI 12 IB-4 g8 & Catalina Island, California, USA\\
	\textit{D. sechellia} & 14021-0248.25 & (Jones) Robertson 3C & Cousin Island, Seychelles \\
	\textit{D. willistoni} & 14030-0811.24 & Powell Gd-H4-1 & Guadeloupe Island, France \\
	\textit{D. ananassae} & 14024-0371.13 & Matsuda (AABBg1) & Hawaii, USA \\
	\hline
\end{tabular}
\label{tab:SpeciesList}	
\end{table}

\begin{table}[!ht]
\caption{\bf{Major morphological events in \textit{Drosophila} development}}
\begin{tabular}{| p{5.2cm} | c | p{7cm} |}

	\hline
	Event  & Stage \cite{Campos-Ortega1985, Bownes1975} & Notes \\
	\hline
	Posterior gap appears & 2 & Gap between yolk and vitelline membrane\\
	\textbf{Pole bud appears} & 3 & \textbf{Cells migrate into the posterior gap}\\
	Nuclei at periphery & 4 & Cells migrate to edges\\
	Pole cells form & 4 & Replication of the pole cells\\
	Yolk contraction & 4 & Light edge of embryo expands\\
	Cellularization begins & 5 & Cell cycle 14\\
	\textbf{Membrane reaches the yolk} & 5 & \textbf{This is regarded as the zero time-point} \\
	Pole cells migrate & 6 & Pole cells begin anterior movement\\ 
	Cephalic furrow forms & 6 & Dorsal and ventral furrows form\\
	\textbf{Pole cells invaginate} & 7 & \textbf{Pole cells enter dorsal furrow}\\
	Transversal fold formation & 7 & Dorsal furrows between amnioproctodeum and cephalic furrow\\
	Cephalic furrow reclines & 8 & Dorsal furrow moves posteriorly \\
	\textbf{Amnioproctodeal invagination} & 8 & \textbf{Invagination approaches cephalic fold}\\
	Anterior midgut primordial & 8 & Tissue thickens at anterior ventral edge\\
	Stomodeal plate forms & 9 & Ventral gap anterior to cephalic fold\\
	Stomodeum invagination & 10 & Ventral furrow anterior to cephalic fold\\
	Clypeolabral lobe forms & 10 & Dorsal, ventral furrows both present\\
	Germ band maxima & 11 & Maximum extension of germband \\
	Clypeolabrum rotates & 11 & Clypeolabrum shifts dorsally\\
	Posterior gap & 11 & Gap forms before germband shortening \\
	Gnathal bud appears & 12 & Ventral tissue between the clypeolabrum and cephalic folds moves anteriorly\\
	Germband retraction begins & 12 & Movement begins mid-germband\\
	\textbf{Amnioserosa exposed} & 12 & \textbf{Germband retracted to the posterior 30\% of the embryo}\\
	Germband retracted & 13 & Germband fully retracted\\
	Dorsal divot & 14 & Dorsal gap between head and amnioserosa\\
	\textbf{Clypeolabrum retracts} & 14 & \textbf{Clypeolabrum pulls away from anterior vitelline membrane}\\
	Anal plate forms & 14 & Posterior depression forms\\
	Midgut unified & 14 & Dark circle forms at embryo's center\\
	\textbf{Heart-shaped midgut} & 15 & \textbf{Triangular midgut}\\
	Clypeolabrum even with ventral lobes & 16 & Ventral lobes move anteriorly to be even with clypeolabrum\\
	Gnathal lobes pinch & 16 & Gnathal lobes meet\\
	Convoluted gut & 16 & Separation between sections of the midgut\\
	Head involution done & 17 & Head lobes complete anterior migration\\
	Muscle contractions & 17 & Head begins twitching\\
	\textbf{Trachea fills} & 17 & \textbf{Developmental end point} \\
	Hatch & 17 & Highly variable \\
	\hline
\end{tabular}
\label{tab:EventList}	
\end{table}

\begin{table}[!ht]
\caption{\bf{\textit{Drosophila} development videos}}
\begin{tabular}{| l | l | }
\hline
Subject & Link \\
\hline
\textit{D. melanogaster} with labelled stages & \url{http://www.youtube.com/watch?v=dYSrXK3o86I} \\
\textit{D. melanogaster} with labelled stages at reduced framerate & \url{http://www.youtube.com/watch?v=QKVmRy3dDR0} \\
\textit{D. melanogaster} at 7 temperatures & \url{http://www.youtube.com/watch?v=-yrs4DcFFF0} \\
11 species at 17.5\degree C & \url{http://www.youtube.com/watch?v=HId_Idz-GhQ} \\
11 species at 22.5\degree C & \url{http://www.youtube.com/watch?v=jO6JfgwMaH4} \\
11 species at 27.5\degree C & \url{http://www.youtube.com/watch?v=vlYeuFqKQhI} \\
\textit{D. ananassae} at 7 temperatures & \url{http://www.youtube.com/watch?v=vy6L4fmWkso} \\
\textit{D. mojavensis}  at 6 temperatures &  \url{http://www.youtube.com/watch?v=XWMs4oUx_mU} \\
\textit{D. virilis} at 6 temperatures & \url{http://www.youtube.com/watch?v=eyr4ckDb0kM} \\
\textit{D. pseudoobscura} at 6 temperatures & \url{http://www.youtube.com/watch?v=sYi-FUXpv4Q} \\

\hline
\end{tabular}
* All videos available at DOI:10.5061/dryad.s0p50
\label{tab:VideoList}
\end{table}

\begin{table}[!ht]
\caption{\bf{The timing of specific developmental events can be predicted as a function of total developmental time}}
\renewcommand{\arraystretch}{1.5}

\begin{tabular}{| l | p{3.5cm} | c |}
\hline
Stage & Event Timing (hours post cellularization) & Percent Error \\
\hline
Pole bud appears & $t_{pba} \approx -0.093t_{dev}$ & 8\%\\
Pole cells invaginate & $t_{pci} \approx 0.018t_{dev}$ & 40\%\\
Amnioproctodeal invagination & $t_{api} \approx 0.035t_{dev}$ & 18\%\\
Amnioserosa exposed & $t_{ase} \approx 0.35t_{dev}$ & 6\%\\
Clypeolabrum retracts & $t_{clr} \approx 0.49t_{dev}$ & 4\%\\
Heart-shaped midgut & $t_{hsm} \approx 0.57t_{dev}$ & 12\%\\
\hline
\end{tabular}
\label{tab:EventTiming}
\end{table}

\begin{table}[!ht]
\caption{\bf{The developmental time of embryos between 17.5\degree C and 27.5\degree C is a species-specific function of temperature}}
\renewcommand{\arraystretch}{1.5}
\begin{tabular}{| l | l | c | p{6cm} | p{1.75cm} |}
\hline
Species & Developmental Time* & $R^{2\dagger} $ & 95\% Confidence Prediction Interval for Future Observations & $Q_{10} ^{\ddagger}$ (27.5:17.5)\\
\hline
\textit{D. virilis}$^{\#}$ & $t_{ Dvir } = 5.64 e^{ 37.08  / T}$ &  0.989 &$t_{ Dvir } \plusminus 31.937 \sqrt{ 1.00 + (\frac{1}{T}- 0.04 )^{2}}$ & 2.2 \\
\textit{D. mojavensis}$^{\#}$ & $t_{ Dmoj } = 3.67 e^{ 43.81  / T}$ &  0.983 &$t_{ Dmoj } \plusminus 54.263 \sqrt{ 1.00 + (\frac{1}{T}- 0.05 )^{2}}$ & 2.5 \\
\textit{D. willistoni} & $t_{ Dwil } = 3.63 e^{ 40.50 / T}$ &  0.944 &$t_{ Dwil } \plusminus 3.122 \sqrt{ 1.00 + (\frac{1}{T}- 0.04 )^{2}}$ & 2.3 \\
\textit{D. pseudoobscura} & $t_{ Dpse } = 7.61 e^{ 25.95 / T}$ &  0.903 &$t_{ Dpse } \plusminus 39.257 \sqrt{ 1.00 + (\frac{1}{T}- 0.05 )^{2}}$ & 1.7 \\
\textit{D. persimilis} & $t_{ Dper } = 9.31 e^{ 21.20 / T}$ &  0.961 &$t_{ Dper } \plusminus 22.598 \sqrt{ 1.00 + (\frac{1}{T}- 0.05 )^{2}}$ & 1.6 \\
\textit{D. ananassae} & $t_{ Dana } = 2.94 e^{ 42.68 / T}$ &  0.979 &$t_{ Dana } \plusminus 1.440 \sqrt{ 1.00 + (\frac{1}{T}- 0.05 )^{2}}$ & 2.4 \\
\textit{D. yakuba} & $t_{ Dyak } = 4.67 e^{ 33.08 / T}$ &  0.943 &$t_{ Dyak } \plusminus 2.203 \sqrt{ 1.00 + (\frac{1}{T}- 0.05 )^{2}}$ & 2.0 \\
\textit{D. erecta} & $t_{ Dere } = 5.21 e^{ 32.97 / T}$ &  0.937 &$t_{ Dere } \plusminus 2.689 \sqrt{ 1.00 + (\frac{1}{T}- 0.04 )^{2}}$ & 2.0 \\
\textit{D. melanogaster} & $t_{ Dmel } = 4.02 e^{ 37.31 / T}$ &  0.963 &$t_{ Dmel } \plusminus 1.281 \sqrt{ 1.00 + (\frac{1}{T}- 0.05 )^{2}}$ & 2.2 \\
\textit{D. sechellia} & $t_{ Dsec } = 4.47 e^{ 34.67 / T}$ &  0.957 &$t_{ Dsec } \plusminus 2.386\sqrt{ 1.00 + (\frac{1}{T}- 0.04 )^{2}}$ & 2.1 \\
\textit{D. simulans} & $t_{ Dsim } = 3.50 e^{ 39.14 / T}$ &  0.960 &$t_{ Dsim } \plusminus 1.883 \sqrt{ 1.00 + (\frac{1}{T}- 0.05 )^{2}}$ & 2.3 \\
\hline
\end{tabular}
\label{tab:SpeciesTiming}
\begin{flushleft}
* End of cellularization to trachea fill in hours, where T is in \degree C

$\dagger R^2$, the Pearson Product-Moment's Correlation Coefficient of determination, is calculated following a least-squares regression across all data points to a curve of the form ln(developmental time) = b($1\over{T}$)+a.

$\ddagger$ $Q_{10}$ is the ratio between developmental times across a 10 degree interval, in this case between 27.5\degree C and 17.5\degree C. A value of 2.2 would indicate that development takes 2.2 times as long at 17.5\degree C than at 27.5\degree C.

$\#$ Curve fit through 30\degree C
\end{flushleft}

\end{table}
\clearpage

\section*{Supplementary Figures}
\beginsupplement

\end{doublespacing}
\end{document}